\documentclass[twocolumn,aps,prb,showpacs,amsmath,superscriptaddress,longbibliography,notitlepage]{revtex4-1}
\usepackage{amssymb}
\usepackage{mathrsfs}
\usepackage{graphicx}
\usepackage{float}
\usepackage[caption=false]{subfig}
\usepackage{array}
\usepackage{epstopdf}

\usepackage[normalem]{ulem}
\usepackage{color}
\usepackage{bm}

\def\CH{\textcolor{black}}
\def\blue{\textcolor{blue}}
\def\red{\textcolor{red}}

\begin{document}

\def\qv{\vec{q}}
\def\red{\textcolor{red}}
\def\blue{\textcolor{black}}
\def\magenta{\textcolor{magenta}}
\def\apricot{\textcolor{Apricot}}

\def\GJ{\textcolor{black}}
\def\YH{\textcolor{black}}
\def\TT{\textcolor{black}}

\newcommand{\norm}[1]{\left\lVert#1\right\rVert}
\newcommand{\ad}[1]{\text{ad}_{S_{#1}(t)}}

\title{Geometric classification of non-Hermitian topological systems through the singularity ring}

\author{Linhu Li}
\affiliation{Department of Physics, National University of Singapore, Singapore 117551, Republic of Singapore}
\author{Ching Hua Lee}
\affiliation{Department of Physics, National University of Singapore, Singapore 117551, Republic of Singapore}
\affiliation{Institute of High Performance Computing, A*STAR, Singapore, 138632.}
\author{Jiangbin Gong}  \email{phygj@nus.edu.sg}
\affiliation{Department of Physics, National University of Singapore, Singapore 117551, Republic of Singapore}


\date{\today}
\begin{abstract}
{This work unveils how geometric features of two-band non-Hermitian Hamiltonians can completely classify the topology of their eigenstates and energy manifolds. Our approach generalizes the Bloch sphere visualization of Hermitian systems to a ``Bloch torus'' picture for non-Hermitian systems, where a singularity ring (SR) captures the degeneracy structure of generic exceptional points.
The SR picture affords convenient visualization of various symmetry constraints and reduces their topological characterization to the classification of simple intersection or winding behavior, as detailed by our explicit study of chiral, sublattice, particle-hole and conjugated particle-hole symmetries. In 1D, the winding number about the SR corresponds to the band vorticity measurable through the Berry phase. In 2D, more complicated winding behavior leads to a variety of phases that illustrates the richness of the interplay between SR topology and geometry beyond mere Chern number classification. Through a normalization procedure that puts generic 2-band non-Hermitian Hamiltonians on equal footing, our SR approach also allows for vivid visualization of the non-Hermitian skin effect.}
\end{abstract}

\maketitle
\section{Introduction}



In recent years, increased attention has been placed on the topological properties of non-Hermitian phases, which behave very differently from Hermitian topological phases of matter~\cite{Lee2016nonH,Hassan2017EP,Hu2017EP,Martinez2018nonH,Lee2018Anatomy,Fu2018nonHermitian,Yin2018nonHermitian,Yao2018nonH2D,Yao2018nonH1D,Kunst2018biorthogonal,Xiong2018BBC,
Longwen2018nonH,Longwen2018nonHFloquet,Hui2018nonH,Gong2018nonHclass,Kawabata2018nonHclass,Liu2019nonHclass,Song2019BBC,Nori2019HOnonH,Edvardsson2019HOnonH,Lee2018hybrid,Ezawa2018HOnonH,Luo2019HOnonH}.
\CH{In non-Hermitian systems, different energy bands can intersect at exceptional points (EPs) where two or more eigenmodes coalesce and form an incomplete Hilbert basis, even when the Hamiltonian remains non-vanishing}~\cite{Hassan2017EP,Hu2017EP,berry2004EP,rotter2009EP,heiss2012EP}. Another fascinating phenomenon is the non-Hermitian skin effect (NHSE), where eigenmodes all accumulate along the boundaries due to non-Hermitian pumping~\cite{Lee2016nonH,Martinez2018nonH,Lee2018Anatomy}, leading to a modification to the conventional bulk-boundary correspondence (BBC)~\cite{Yao2018nonH1D,Kunst2018biorthogonal,Xiong2018BBC,Song2019BBC}.
Besides these intriguing non-Hermitian behaviors, many novel topological properties of Hermitian systems have also been extended and widely studied in the context of non-Hermiticity, including the newly emerging higher-order topological phases~\cite{Nori2019HOnonH,Edvardsson2019HOnonH,Lee2018hybrid,Ezawa2018HOnonH,Luo2019HOnonH,benalcazar2017HOTI,Benalcazar2017HOTI2},
{and the long-term quest towards symmetry classification of topological phases~\cite{Gong2018nonHclass,Kawabata2018nonHclass,Liu2019nonHclass,schnyder2008class,ryu2010class}.}


In Hermitian systems, topological properties can be most easily visualized through the geometric behavior of the Hamiltonian in the pseudospin space. For instance, one-dimensional (1D) chiral-symmetry protected topological systems can be {characterized by a $Z$ winding number that represents the total number of times the path of the configuration vector of the Hamiltonian encircles} the origin of the pseudospin space~\cite{Niu2012Majorana,Song2015Ising,Li2015winding}. The quantized Berry phase for 1D $Z_2$ systems is equivalent to the solid angle of the winding path, whose quantization is protected by particle-hole symmetry~\cite{Li2016EPJB}. In two-component two-dimensional (2D) systems, the winding path forms a 2D manifold in a 3D pseudospin-1/2 space, such that Chern number corresponds to the total number of times that manifold encloses the origin~\cite{asboth2016short}, or the number of poles when mapped to the complex projective plane~\cite{lee2017band}.
However, a general geometric interpretation of non-Hermitian Hamiltonians is still lacking in the literature (except for a specific chiral-symmetry-protected case~\cite{Yin2018nonHermitian}). {Developing it will provide new insights into the interplay between the geometric and topological properties of generic non-Hermitian Hamiltonians, and shed light on their mathematical links with existing Hermitian systems with highly nontrivial topological loops~\cite{bode2016constructing,bi2017nodal,chang2017weyl,ezawa2017topological,li2018realistic,yan2018experimental,gao2018experimental,luo2018topological,lee2019imaging}.}


{In this paper, we geometrically study two-component non-Hermitian Hamiltonians and provide unifying pictures with regards to conventional properties like the Berry phase and Chern number, as well as non-Hermitian constructs like vorticity and the skin effect. With a three-dimensional (3D) pseudospin vector space, the degeneracy structure of non-Hermitian exceptional points effectively extends the origin} from a zero-dimensional (0D) point to a 1D ring which we call the singularity ring (SR). Compared with the 0D origin, the 1D SR possesses richer geometric interplay with the winding path of the Hermitian part of the Hamiltonian.
We find that in 1D systems, the linkage between the 1D winding path and the SR 
corresponds to the vorticity, a topological invariant with no Hermitian analogy~\cite{Fu2018nonHermitian}. This linkage corresponds to the $\pi$ Berry phase when one eigenmode returns to itself after two periods~\cite{Mailybaev2005geometric}, and can alternatively be mapped to the Chern number of an effective 2D Chern insulator. 
{In generic cases where the non-Hermiticity takes on complicated momentum dependence, the SR can still be mapped back into a unit ring via a basis normalization and rotation. This allows access to geometric visualizations of the NHSE, where effect of the equivalent non-Bloch basis rescaling is made intuitive.} As such, the SR approach proves useful for studying the spectra under both \CH{periodic boundary conditions (PBC)} and {open boundary conditions (OBCs)}, {thus directly relating topological boundary modes with geometric features of the Hamiltonian. 
Furthermore, the SR picture can be generalized to higher dimensions, where the winding around the SR assumes fancier manifolds like the torus.}



The rest of the paper is organized as follows. In Section~\ref{sec:SR}, we introduce and define the SR, and show that its linking number is equivalent to {an associated Chern number, which can be measured} via a summed Berry phase over the two bands. In Section~\ref{sec:topology} we address the NHSE and characterize topological boundary modes from the SR perspective {through detailed studies of four different symmetry constraints.} Next, we extend our geometric interpretation of the SR to 2D non-Hermitian systems in Section~\ref{sec:2D}, {where various new topological and geometric scenarios are unveiled.} Finally, all results are briefly summarized in Section~\ref{sec:sum}.

\section{Singularity ring in 1D two-component non-Hermitian systems}\label{sec:SR}

\subsection{General definition of the singularity ring}

We consider a general 1D two band model described by the Hamiltonian
\begin{eqnarray}
H(k)=\sum_{n=x,y,z}h_n\sigma_n=\sum_{n=x,y,z}(d_n(k)+ig_n)\sigma_n,\label{H_general}
\end{eqnarray}
with $d_n(k)$ and $g_n$ being real, and $\sigma_n$ the Pauli matrices acting on a pseudospin-1/2 space. {To most directly illustrate the SR geometry, we shall first assume constant non-Hermitian terms $g_{x,y,z}$. The more generic case of $k$-dependent non-Hermitian terms $g_{x,y,z}(k)$ will be treated via a normalization procedure as discussed later.} In Hermitian cases with ${\bm g}=(g_x,g_y,g_z)=0$, ${\bm h}(k)=[h_x(k),h_y(k),h_z(k)]$ describes a 3D pseudospin vector, whose winding on the Bloch sphere corresponds to topological properties of the system. However, such a geometric interpretation cannot be directly mapped to non-Hermitian systems as ${\bm h}(k)$ becomes complex. Nevertheless, \CH{a hint arises from the form of the eigen-energies} of the non-Hermitian system of Eq. (\ref{H_general}), which are given by
\begin{eqnarray}
E_{\pm}=\pm\sqrt{\sum_{n=x,y,z} (d_n^2(k)-g_n^2-2ig_nd_n(k))},
\end{eqnarray}
\CH{and vanishing at}
\begin{eqnarray}
&&d_x^2(k)+d_y^2(k)+d_z^2(k)=g_x^2+g_y^2+g_z^2,~{\rm and}\label{con1}\\
&&g_xd_x(k)+g_yd_y(k)+g_zd_z(k)=0\label{con2}.
\end{eqnarray}
These degeneracy conditions {describe} a 1D {ring} in the 3D vector space of $[d_x(k),d_y(k),d_z(k)]$ for each individual $k$, with a radius of $\sqrt{g_x^2+g_y^2+g_z^2}$ and a normal vector given by ${\bm g}=(g_x,g_y,g_z)$.
{Given that the set of ${\bm k}$ defines where the two bands of the Hamiltonian Eq. (\ref{H_general}) become degenerate, we shall refer to this ring of EPs as the singularity ring (SR) hereafter.} Such a geometric interpretation can be viewed as an {\emph{extension of the Bloch sphere}} for Hermitian systems, with the origin of pseudospin space {generalized to the whole SR},
and the pseudospin vector replaced by the ${\bm d}(k)$ vector, whose winding  gives a 1D closed loop as $k$  varies from $0$ to $2\pi$.

\begin{figure}
\includegraphics[width=0.9\linewidth]{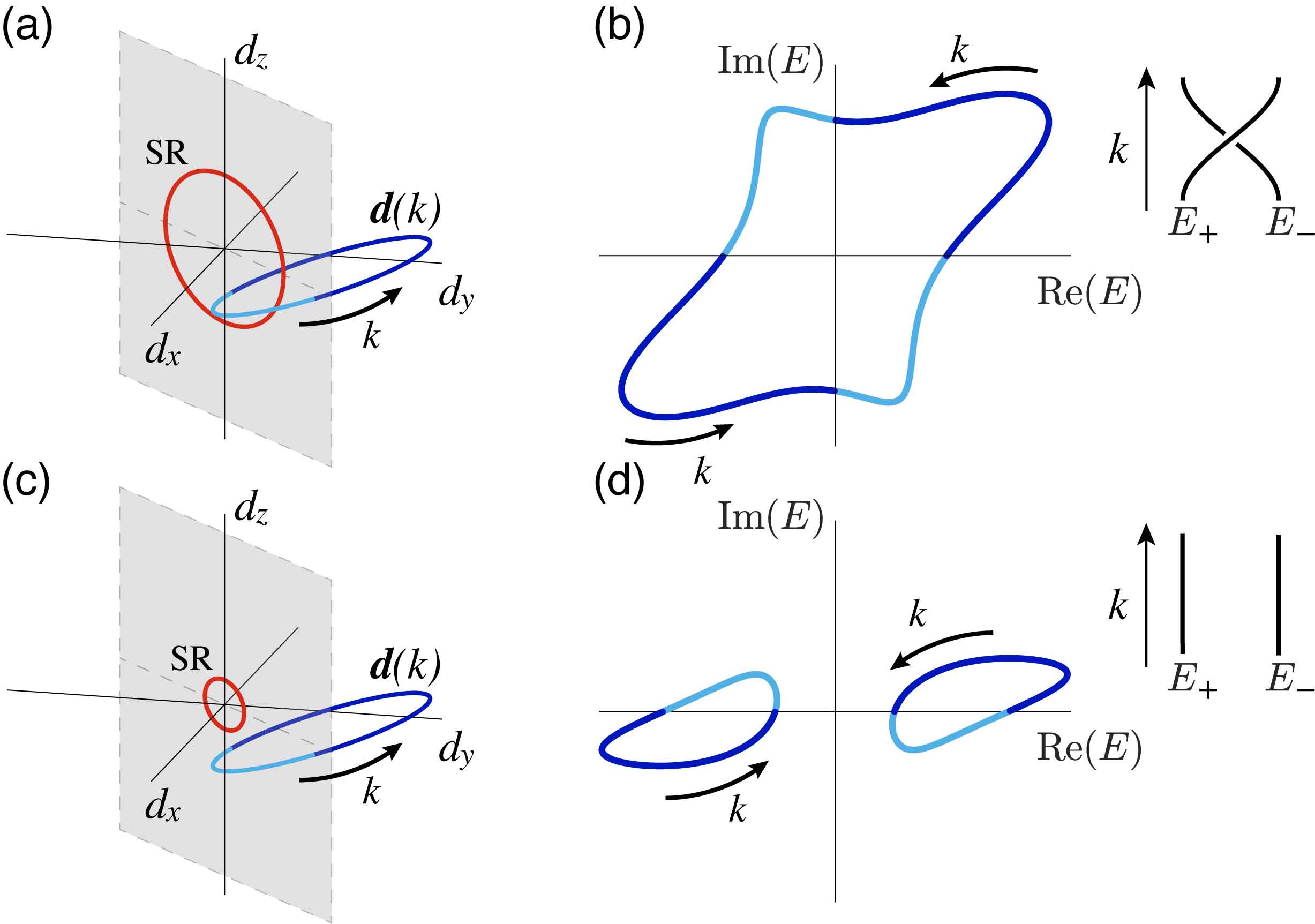}
\caption{Sketches of different linking relations between the SR and the ${\bm d}(k)$-loop, and their corresponding complex spectra. (a,c) The red circle indicates the SR, and the blue trajectories show the ${\bm d}(k)$-loop, with lighter and darker colors indicating the parts on different sides of the plane containing the SR respectively. (b,d) The spectrum of $H(k)$ corresponding to the regions with the same colors in (a,c) respectively. }
\label{SR_sketch}
\end{figure}

\CH{The linking of the SR with the ${\bm d}(k)$-loop has a direct interpretation in terms of the vorticity~\cite{Fu2018nonHermitian} of the bands. To see this, the}  eigen-energies $E_\pm$ fall \CH{into} different quadrants in the complex plane for ${\bm d}(k)$ on \blue{different sides of the SR-plane [plane containing the SR, as given by Eq. (\ref{con2})]}, and take purely imaginary/real values when ${\bm d}(k)$ intersects the SR-plane inside/outside the SR. Thus the two bands are \CH{connected as} one closed trajectory \CH{i.e. have half-integer vorticity} when the SR and the ${\bm d(k)}$-loop are linked, as shown in Fig.~\ref{SR_sketch}(a,b).
This joint spectrum corresponds to a braiding of the two bands along $k$, as shown by the inset of Fig.~\ref{SR_sketch}(b).

In Fig.~\ref{SR_sketch}(c) and (d), we illustrate another case where the ${\bm d}(k)$-loop and SR are unlinked, where the two bands are separated from each other.
In general, {periodic boundary condition (PBC)} spectra corresponding to different linkage cannot be transformed into each other without going through {the band-touching SR}, leading to a topological classification of the linking number. This linking number reflects the geometric {property} of the vorticity of the two bands, which {is a unique topological property of} the non-Hermitian systems considered here.

\subsection{\CH{SR normalization for} $k$-dependent non-Hermitian terms}\label{sec:normalization}
In above discussion, we have only considered $k$-independent non-Hermitian terms, i.e. a constant ${\bm g}$ vector. More generally, a $k$-dependent \CH{non-Hermitian terms} ${\bm g}(k)$ gives a SR with its radial size and orientation also being $k$-dependent. {To preserve a meaningful notion of SR linkage, we will need to rotate and normalize the SR for each $k$ onto the same unit ring}, without changing its topological relation to the ${\bm d}(k)$-loop.
{Without loss of generality},
we set $(g_x',g_y',g_z')=(0,0,1)$ \CH{as the non-Hermitian terms after rotation and normalization. This can be obtained by defining a new basis, such that the original Hamiltonian $h$ takes the form}
\begin{eqnarray}
H'=\mathcal{R}_y(-\psi)\mathcal{R}_z(-\theta)\frac{H}{r_2}\mathcal{R}^{\dagger}_z(-\theta)\mathcal{R}^{\dagger}_y(-\psi),\label{normalization}
\end{eqnarray}
with $\cos\theta=g_x/r_1$, $\cos\psi=g_z/r_2$, $r_1=\sqrt{g_x^2+g_y^2}$, $r_2=\sqrt{g_x^2+g_y^2+g_z^2}$, and $\mathcal{R}_n(\theta)=e^{-i\frac{\theta}{2}\sigma_n}$ the operator rotating the system around $n$ axis for an angle of $\theta$. The rotation \CH{matrices} satisfy
\begin{eqnarray}
\mathcal{R}_z(\theta)\sigma_x\mathcal{R}^{\dagger}_z(\theta)&=&\sigma_x\cos\theta+\sigma_y\sin\theta\nonumber\\
\mathcal{R}_z(\theta)\sigma_y\mathcal{R}^{\dagger}_z(\theta)&=&\sigma_y\cos\theta-\sigma_x\sin\theta\nonumber\\
\mathcal{R}_z(\theta)\sigma_z\mathcal{R}^{\dagger}_z(\theta)&=&\sigma_z\nonumber,
\end{eqnarray}
\CH{with the actions of $\mathcal{R}_x(\theta)$ and $\mathcal{R}_y(\theta)$ given by permuting the above via} $x\rightarrow y\rightarrow z\rightarrow x$. \CH{Explicitly, $H'$ takes the form}
\begin{eqnarray}
r_2H'&=&[(d_x\cos\theta+d_y\sin\theta)\cos\psi-d_z\sin\psi]\sigma_x\nonumber\\
&&+[(d_x\cos\theta+d_y\sin\theta)\sin\psi+d_z\cos\psi+r_2i]\sigma_z\nonumber\\
&&+(d_y\cos\theta-d_x\sin\theta)\sigma_y,
\end{eqnarray}
\CH{indeed possessing a constant non-Hermitian term $i\sigma_z$ giving rise to a SR which is a unit ring in the $x-y$ plane. Through this normalization, generic two-component non-Hermitian Hamiltonians can always be rewritten as Hamiltonians with constant unit non-Hermitian term. Although we will only study physical Hamiltonians with constant non-Hermitian terms in this paper, this normalization procedure will still be necessary for the 1D and 2D systems with SLS and PHS$^{\dagger}$ symmetry, which sees effective $k$-dependent non-Hermitian terms induced by the non-Bloch basis.}

\subsection{The linking number and \CH{the} Chern number}
After the \CH{normalization and rotation} procedure in Sec.~\ref{sec:normalization}, the linking number can be directly counted \CH{geometrically}, as in Fig.~\ref{SR_sketch}(a,c). Alternatively, \CH{this linking number can also be obtained by computing the Chern number of an associated 2D system, without having to normalize the SR for each $k$.} 
{To construct the 2D description, we first parameterize the original (un-normalized)} SR with a phase parameter $\delta$
{along its circumference:}
\begin{eqnarray}
h_{{\rm SR},x}(k,\delta)&=&r(\tilde{u}_x\cos \delta+\tilde{v}_x\sin \delta),\nonumber\\
h_{{\rm SR},y}(k,\delta)&=&r(\tilde{u}_y\cos \delta+\tilde{v}_y\sin \delta),\nonumber\\
h_{{\rm SR},z}(k,\delta)&=&r(\tilde{u}_z\cos \delta+\tilde{v}_z\sin \delta),
\end{eqnarray}
with {$k$-dependence contained in }$r=\sqrt{g_x^2+g_y^2+g_z^2}$ the radius of the SR, ${\bm n}=(g_x,g_y,g_z)$ the normal vector, ${\bm u}=(g_y,-g_x,0)$ a vector lying in the same plane with the SR, ${\bm v}={\bm n} \times {\bm u}$, and $\tilde{\bm u}$ and $\tilde{\bm v}$, the normalized vectors of ${\bm u}$ and ${\bm v}$.
Next we consider the \CH{winding properties of the }vector
\begin{eqnarray}
{\bm h}_{\rm 2D}(k,\delta)={\bm d}(k)-{\bm h}_{\rm SR}(k,\delta),
\end{eqnarray}
which depends on two parameters $k$ and $\delta$. This vector describes a 2D manifold in the 3D vector space, which can be {visualized as the trajectory of ${\bm d}(k)$ shifted by ${\bm h}_{\rm SR}(k,\delta)$} for each $\delta$. This procedure also shifts each point on the SR to the origin, and the linking number of the SR and the ${\bm d}(k)$-loop \CH{reduces to} how many times that the 2D manifold \CH{traced out by} ${\bm h}_{\rm 2D}(k,\delta)$ encloses the origin.
Therefore the Chern number $C_L$ of an {associated} 2D Hamiltonian
\begin{eqnarray}
H_{\rm 2D}(k,\delta)={\bm h}_{\rm 2D}(k,\delta)\cdot {\bm \sigma},\label{effective_2D}
\end{eqnarray}
 gives the linking number of the SR and the ${\bm d}(k)$-loop. In other words, a 1D two-component non-Hermitian system {has been mapped to a 2D Hermitian system, with the new dimension reflecting the intrinsic 1D nature of the singularity structure. This approach complements the above-mentioned normalization procedure by relating geometric winding around the normalized ring to the 2nd homotopy around an associated Bloch sphere.}

\subsection{The linking number and \CH{the} Berry phase}\label{sec:SR_Berry}
\begin{figure}
\includegraphics[width=0.99\linewidth]{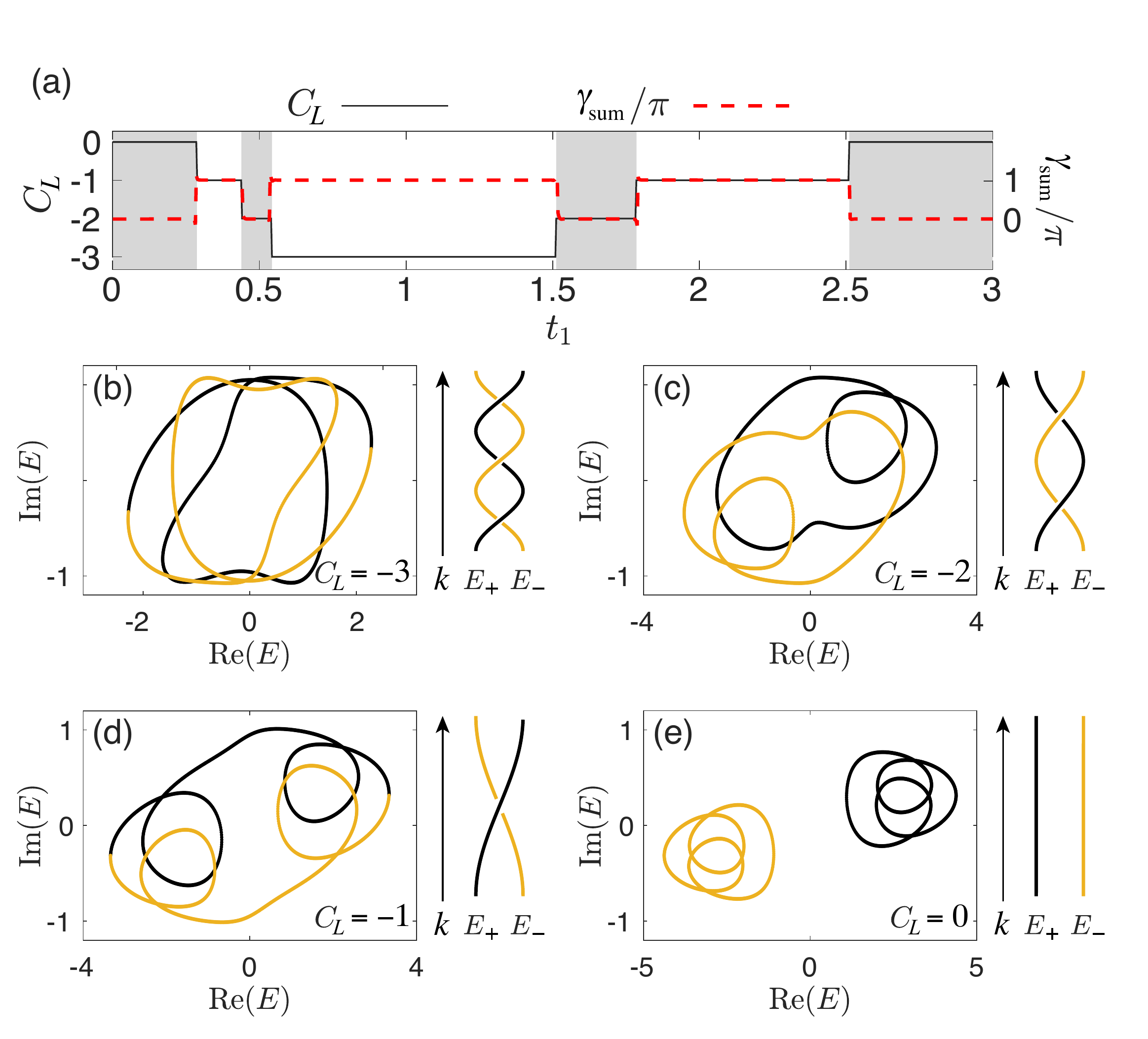}
\caption{(a)The linking number $C_L$ as a function of $t_1$ for the Hamiltonian (\ref{eSSH}).
(b-e) Complex spectra (with yellow and black indicating the two bands) and sketches of the braiding structures of the system with $t_1=1$, $1.7$, $2$, and $3$.
Other parameters are $t_2=0.5$, $t'=1$, $g_x=g_z=0.3$, $g_y=1$.
}
\label{evolution_eSSH}
\end{figure}

In this subsection, we establish the connection between {a linkage with the SR} and a Berry phase defined for both bands of the system. Since an exceptional point shall emerge when the  ${\bm d}(k)$-loop touches the SR,
a nontrivial linkage between the SR and the ${\bm d}(k)$-loop indicates that the system goes around an exceptional degeneracy when $k$ varies from $0$ to $2\pi$.
Therefore, {the system is expected to acquire a quantized $\pi$ Berry phase when $\bold d(k)$ goes through the Brillouin zone twice~\cite{Mailybaev2005geometric}.
More specifically, since the two bands are now connected as one closed loop, this process of cycling around the EP necessarily involves both bands}. The total Berry phase of the two bands should be quantized to $\pi$. The total Berry phase can be obtained from the sum of the Berry phases of the two bands, i.e.,
\begin{eqnarray}
\gamma_{\rm sum}=\gamma_- +\gamma_+,\label{total_phase}
\end{eqnarray}
with\cite{zhang2019QGT}
\begin{eqnarray}
\gamma_\pm=-{\rm Im}\oint_0^{2\pi}dk \frac{\langle u_\pm^L(k)|\partial_k|u_\pm^R(k)\rangle}{\langle u_\pm^L(k)|u_\pm^R(k)\rangle},\label{Berry_phase}
\end{eqnarray}
$u_\pm^L(k)$ and $u_\pm^R(k)$ being the left and right Bloch state eigenvectors corresponding to $E_\pm$.
{That is, $\gamma_{\rm sum}=\pi$ when the two bands form a single closed loop, and $\gamma_{\rm sum}=0$ otherwise i.e.
\begin{equation}
\gamma_{\rm sum}=\pi\,(C_L\text{ mod }2)
\end{equation}
}
To give a concrete example, we consider an extended version of the Su-Schrieffer-Heeger (SSH) model~\cite{SSH},
\begin{eqnarray}
h(k)&=&(t_1+t_2\cos{k}+t'\cos{3k}+ig_x)\sigma_x\nonumber\\
&&+(t_2\sin{k}+t'\sin{3k}+ig_y)\sigma_y+ig_z\sigma_z,\nonumber\\
\label{eSSH}
\end{eqnarray}
with $t'$ \CH{giving rise to a linking number of up to $\pm3$}. We numerically calculate the Berry phase $\gamma_{\rm sum}$ [red dashed line in Fig.~\ref{evolution_eSSH}(a)], and compare it with the linking number (Chern number) $C_L$ [black line Fig.~\ref{evolution_eSSH}(a)]. We can see that $\gamma_{\rm sum}=\pi~(0)$ when $C_L$ takes odd (even) values, i.e. the real winding path encloses the SR an odd (even) number of times. {The corresponding complex spectra of these case with different linking numbers are shown in Fig.~\ref{evolution_eSSH}(b-e)}.
{When $C_L=-2$ for instance, the two bands braid around each other twice and remain disjoint, hence corresponding to $\gamma_{\rm sum}=0$ like the unlinked case with $C_L=0$. As compared to $C_L$ which reflects the \emph{topology} of the associated 2D Hamiltonian, the Berry phase $\gamma_{\rm sum}$ reflects the \emph{geometric} aspects of the eigenvectors, and can be extracted from an adiabatic dynamical process} as discussed in Appendix~\ref{app:dynamical}.

\section{Topological classification from the SR perspective}\label{sec:topology}
{Having discussed how linkages with the SR correspond to nontrivial braidings of the eigenenergies, we shall now show how the SR also provides geometrical interpretations of the \emph{eigenstate} topology. Specifically, the SR picture allows us to visualize how different symmetries lead to different topological classifications in non-Hermitian systems.}

While Hermitian systems can be classified by the time-reversal, particle-hole, and chiral symmetries (TRS, PHS, CS), {each of them possesses a conjugated variant in the non-Hermitian case, leading to a richer classification~\cite{Kawabata2018nonHclass}.}
As the time-reversal symmetry cannot protect 1D topological properties by itself even in Hermitian systems~\cite{schnyder2008class,ryu2010class},
we shall consider only the \blue{other two symmetries and their variants} given by (based on the notation of Ref.~\onlinecite{Kawabata2018nonHclass}):
\begin{eqnarray}
{\rm CS}:&~&\Gamma^{-1}H^{\dagger}(k)\Gamma=-H(k);\nonumber\\
{\rm SLS}:&~&\mathcal{S}^{-1}H(k)\mathcal{S}=-H(k),\nonumber\\
{\rm PHS}^{\dagger}:&~&\mathcal{T}_-^{-1}H^*(k)\mathcal{T}_-=-H(-k);\nonumber\\
{\rm PHS}:&~&\mathcal{C}_-^{-1}H^T(k)\mathcal{C}_-=-H(-k);\nonumber
\end{eqnarray}
where $\Gamma$, $\mathcal{S}$, $\mathcal{T}_-$, and $\mathcal{C}_-$ represent some unitary operators, \blue{whose actual forms depend on the basis and the explicit physical systems under consideration.}
The abbreviation SLS refers to ``sublattice symmetry", which coincides with CS in the presence of Hermiticity.
In Table~\ref{table1}, we summarize some general results.
\CH{The CS and SLS symmetries guarantee that the ${\bm d(k)}$-loop lies in a 2D plane, thereby leading to $Z$-type topological classifications corresponding respectively} to the number of times the ${\bm d(k)}$-loop encloses the SR, or the two intersecting points of the SR and the 2D plane. On the other hand, the PHS$^\dagger$ and PHS symmetries only require a self-symmetric ${\bm d(k)}$-loop. In \CH{these cases, the
systems possess} $Z_2$ topologies related only to the high-symmetric points for PHS$^\dagger$ systems, or to the singularity points of ${\bm d(k)}$-loop and the SR-plane for PHS systems.  {Each of these symmetries are analyzed in detail in the following subsections.}

\begin{table}
\centering
{
\begin{tabular}{|c|c|m{1.7cm}<{\centering}|c|}\hline
Symmetry & NHSE & Topological  quantity  & Geometric features\\    \hline
CS & No & $Z$ &\multicolumn{1}{|m{4cm}|}{\includegraphics[width=0.9\linewidth]{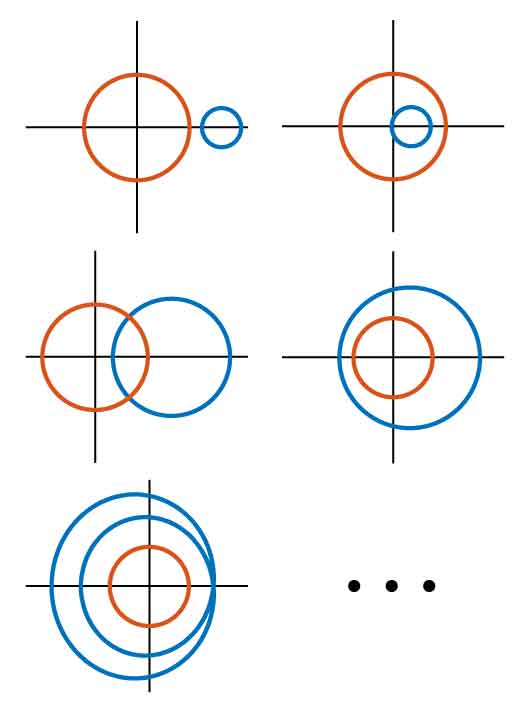}}\\    \hline
SLS & Yes & $Z$ & \multicolumn{1}{|m{4cm}|}{\includegraphics[width=0.9\linewidth]{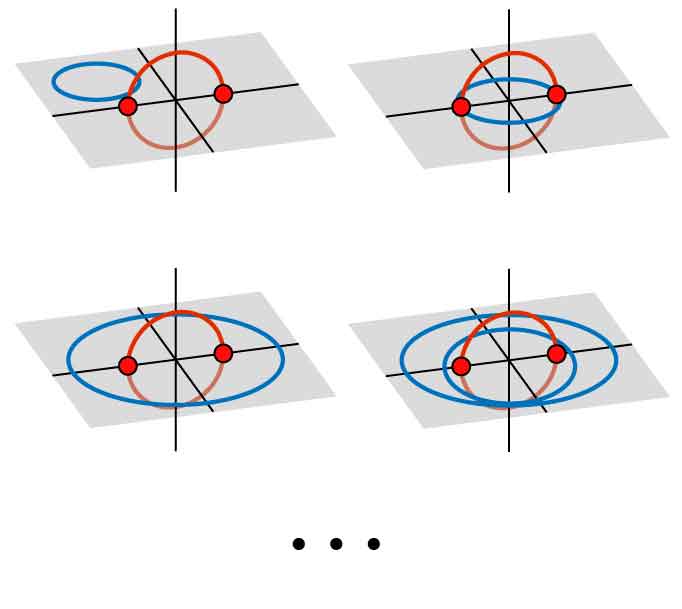}}\\    \hline
PHS$^{\dagger}$ & Yes & $Z_2$ & \multicolumn{1}{|m{4cm}|}{\includegraphics[width=0.9\linewidth]{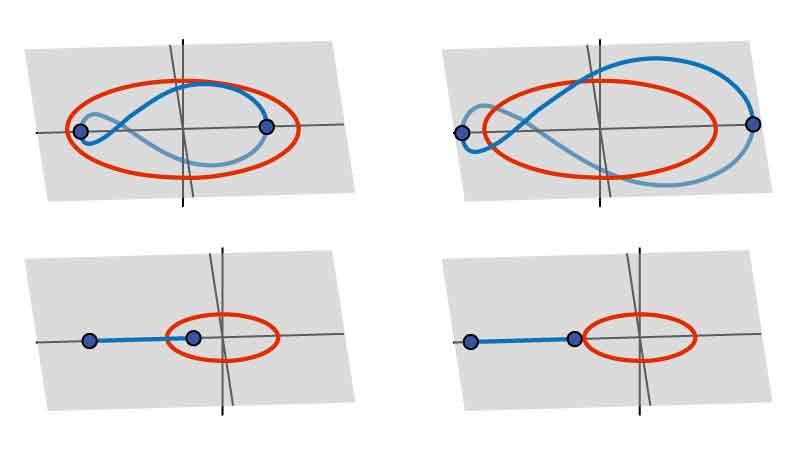}}\\    \hline
PHS & No & $Z_2$ & \multicolumn{1}{|m{4cm}|}{\includegraphics[width=0.9\linewidth]{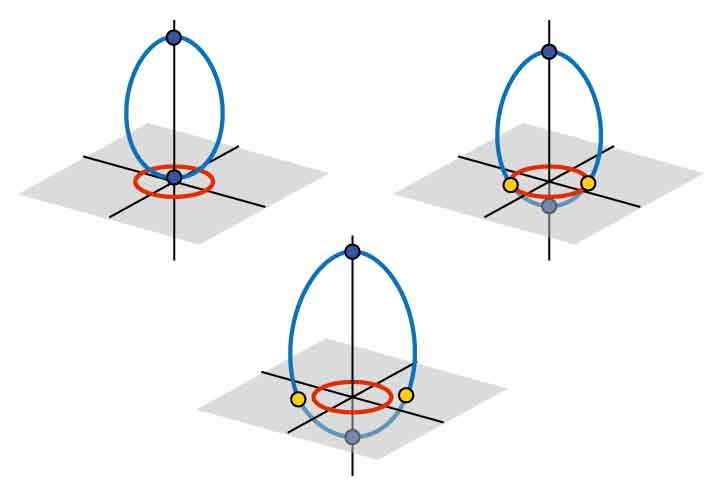}}\\    \hline
\end{tabular}
}\caption{\CH{Summary of the results of systems with different symmetries. In the figures, red and blue loop indicate possible distinct configurations of} the SR and the ${\bm d(k)}$-loop respectively. Red, blue and yellow dots indicate the crucial points related to the topological characterizations, as explained in details in  Sections~\ref{sec:CS} to~\ref{sec:PHS}.}
\label{table1}
\end{table}

\begin{figure}
\includegraphics[width=1\linewidth]{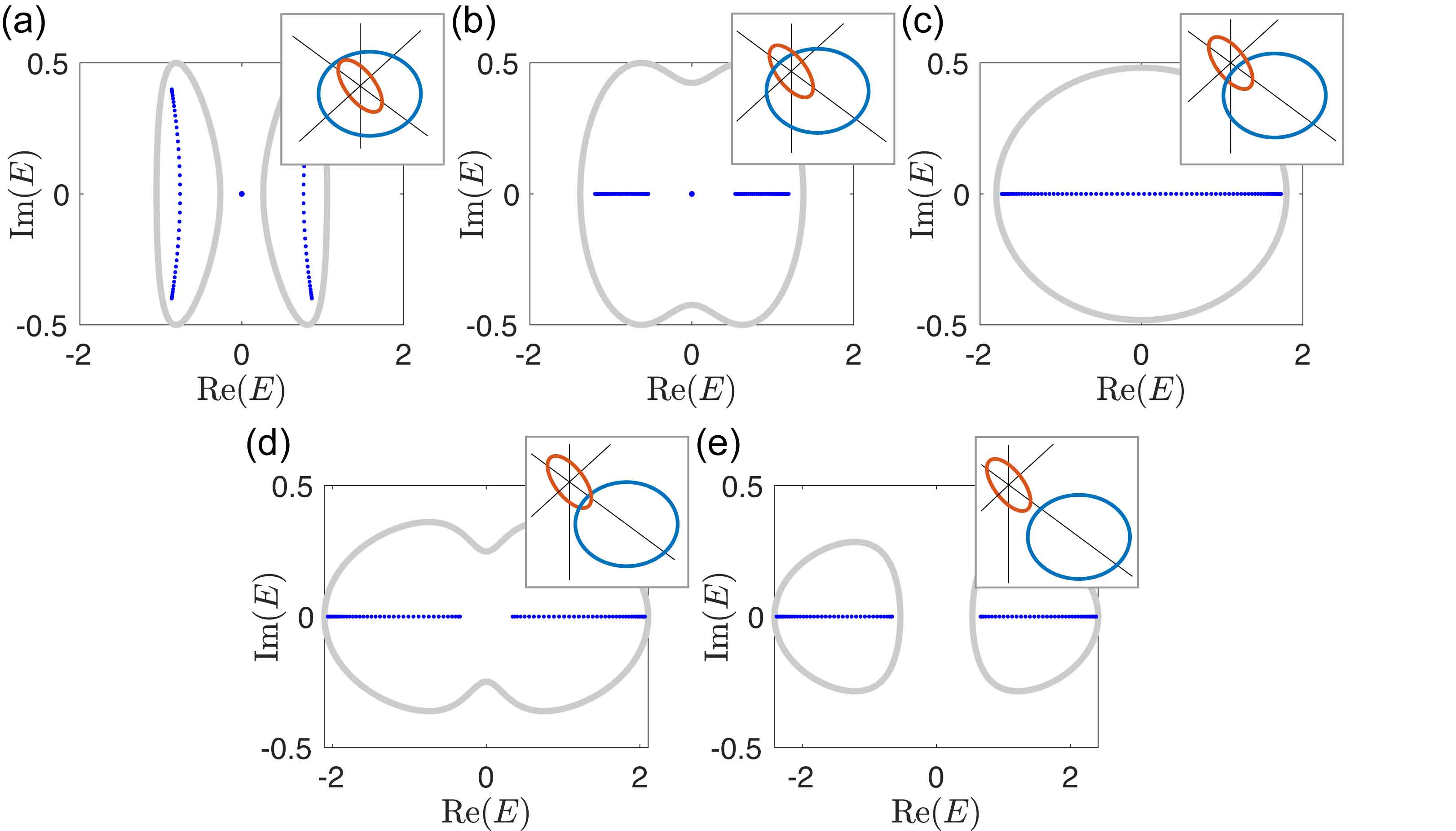}
\caption{Spectrum of Hamiltonian (\ref{SSH_type0}). Gray and dark blue colors indicate the PBC and OBC spectra respectively. The parameters are $g_y=0.5$, $t_2=\sqrt{3}2$, with (a) $t_1=0.3$; (b) $t_1=0.6$; (c) $t_1=1$; (d) $t_1=1.3$; and (e) $t_1=1.6$. The insets show the geometric features, with red and blue loops indicating the SR and the ${\bm d(k)}$-loop.}
\label{spectrum_SSH_type0}
\end{figure}

{In matching the topological boundary states with bulk topological properties of non-Hermitian systems, one caveat is that non-reciprocity may pump all bulk eigenstates towards the boundaries, leading to non-Bloch eigenstates known as skin states. This effect, dubbed as the ``non-Hermitian skin effect'' (NHSE), leads to totally different OBC and PBC eigenspectra, not just for the boundary eigenstates but for all eigenstates~\cite{Yao2018nonH1D,Kunst2018biorthogonal,Xiong2018BBC,Lee2018Anatomy,Song2019BBC}. As such,} the band touchings and topological transitions of the OBC spectrum are not expected to coincide with the PBC band touching points, i.e. \CH{jumps in the linkage between the SR and the ${\bm d(k)}$-loop. Nevertheless,
the OBC spectrum must lie in the interior of the PBC loop in the presence of the NHSE.~\cite{Lee2018Anatomy}.}
Therefore, the band touching of the OBC spectrum (and hence a topological transition) can occur only when the two PBC bands \CH{form a single closed loop, or when they touch}, i.e. when the SR and the ${\bm d(k)}$-loop are linked to each other~\cite{Lee2018Tidal}. We illustrate this constraint on topological phase transition in Fig.~\ref{spectrum_SSH_type0}, which features the generalized non-Hermitian SSH model~\cite{Yao2018nonH1D,Lee2018Anatomy}
\begin{eqnarray}
h(k)=(t_1+t_2\cos{k})\sigma_x+(t_2\sin{k}+ig_y)\sigma_y.\label{SSH_type0}
\end{eqnarray}
{The OBC skin bands (dark blue) are contained within the PBC loops (gray), and are able to touch and annihilate the topological zero mode at the origin only at nonzero vorticity Fig.~\ref{spectrum_SSH_type0}(b,c,d), where the SR and the ${\bm d(k)}$-loop are linked to each other in shown in the insets. Indeed, this topological phase transition occurs at Fig.~\ref{spectrum_SSH_type0}(c), when $t_1^2=g_y^2+t_2^2$. }

\begin{figure}
\includegraphics[width=1\linewidth]{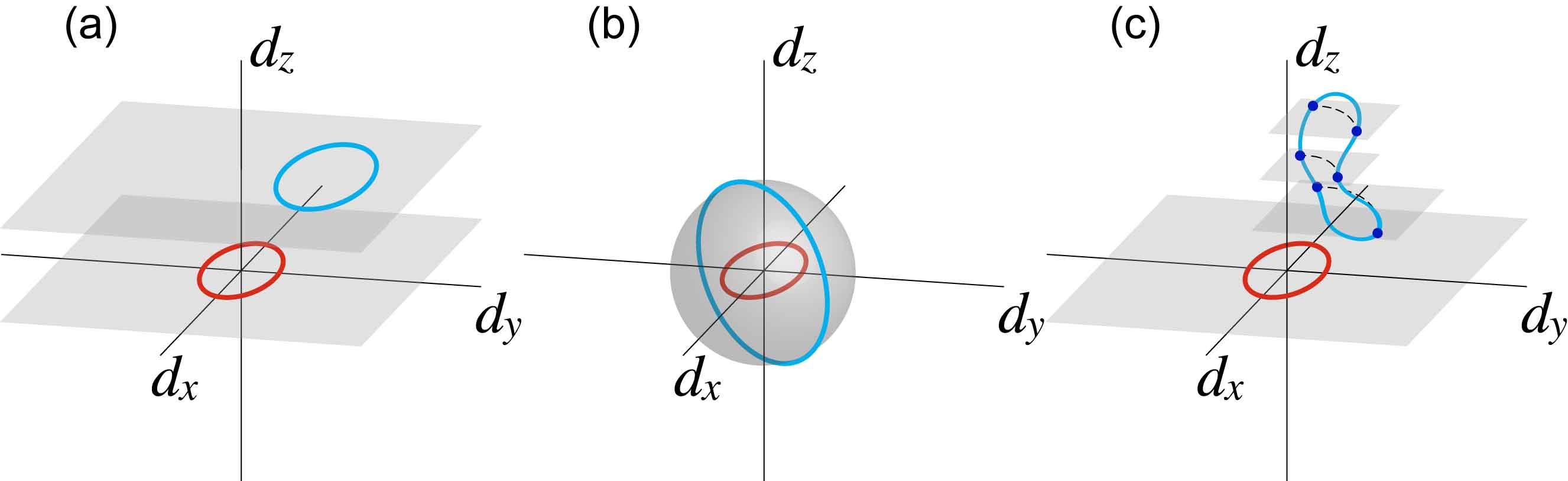}
\caption{\CH{The three scenarios where the NHSE is absent. (a) ${\bm d(k)}$-loop (blue) lies in a plane parallel to the SR (red); (b) ${\bm d(k)}$-loop lies on a sphere center} at the origin; (c) each pair of $k$, e.g. the pairs of blue points, satisfies conditions Eqs.~(\ref{no_skin_con1}) and (\ref{no_skin_con2}). \CH{Scenarios (a) and (b) are special cases of (c).}}
\label{sketch_skin}
\end{figure}

{To geometrically study the topological characteristics of a system, then, it will be crucial to first understand how to detect the NHSE geometrically. A Hamiltonian $h$ without NHSE is reciprocal (satisfies $h(k)=h^T(-k)$)} and have identical PBC and OBC arc-like spectra (except for edge states). 
Therefore its eigenenergies must move along an arc-like spectrum back and forth when $k$ varies from zero to $2\pi$ \cite{Lee2018Anatomy}, leading to at least two-fold degeneracy of a single band, i.e. for any $k_1$, there exists $k_2\neq k_1$, such that
 $E_\pm(k_1)= E_\pm(k_2)$. This condition suggests that both
\begin{eqnarray}
(i)\qquad\sum_{n=x,y,z}d_n(k_{1})g_n=\sum_{n=x,y,z}d_n(k_{2})g_n,\label{no_skin_con1}
\end{eqnarray}
i.e. the line connecting the two points ${\bm d(k_1)}$ and ${\bm d(k_2)}$ is parallel to the SR (in the plane of $\sum_{n=x,y,z}g_n d_n=0$), and
\begin{eqnarray}
(ii)\qquad\sum_{n=x,y,z}d^2_n(k_{1})=\sum_{n=x,y,z}d^2_n(k_{2}),\label{no_skin_con2}
\end{eqnarray}
i.e. the two points have the same modular length.
Therefore, a system exhibits no NHSE only when the ${\bm d(k)}$-loop stays in a plane parallel to the SR [Fig.~\ref{sketch_skin}(a)], or on a sphere whose centering at the origin [Fig.~\ref{sketch_skin}(b)] or, \CH{in a more complicated general scenario, each pair of $k$ points separately satisfies the above conditions [Fig.~\ref{sketch_skin}(c)]. An additional useful fact is that Hamiltonians without NHSE cannot exhibit a linkage between ${\bm d(k)}$ and the SR, because having the linkage will imply the existence of non-degenerate spectral loops (see Fig.~1).}

\CH{As detailed below, one can thus show that of the four symmetries considered, the SLS and \CH{PHS$^\dagger$} cases suffer from the NHSE. Their geometric SR pictures can be constructed after switching to a non-Bloch basis via analytically continuing $k\rightarrow k+i\kappa$, such that the bulk-boundary correspondence is restored~\cite{Yao2018nonH1D,Yao2018nonH2D,Lee2018Anatomy}.}





\subsection{Chiral symmetry (CS)}\label{sec:CS}
\begin{figure*}
\includegraphics[width=1\linewidth]{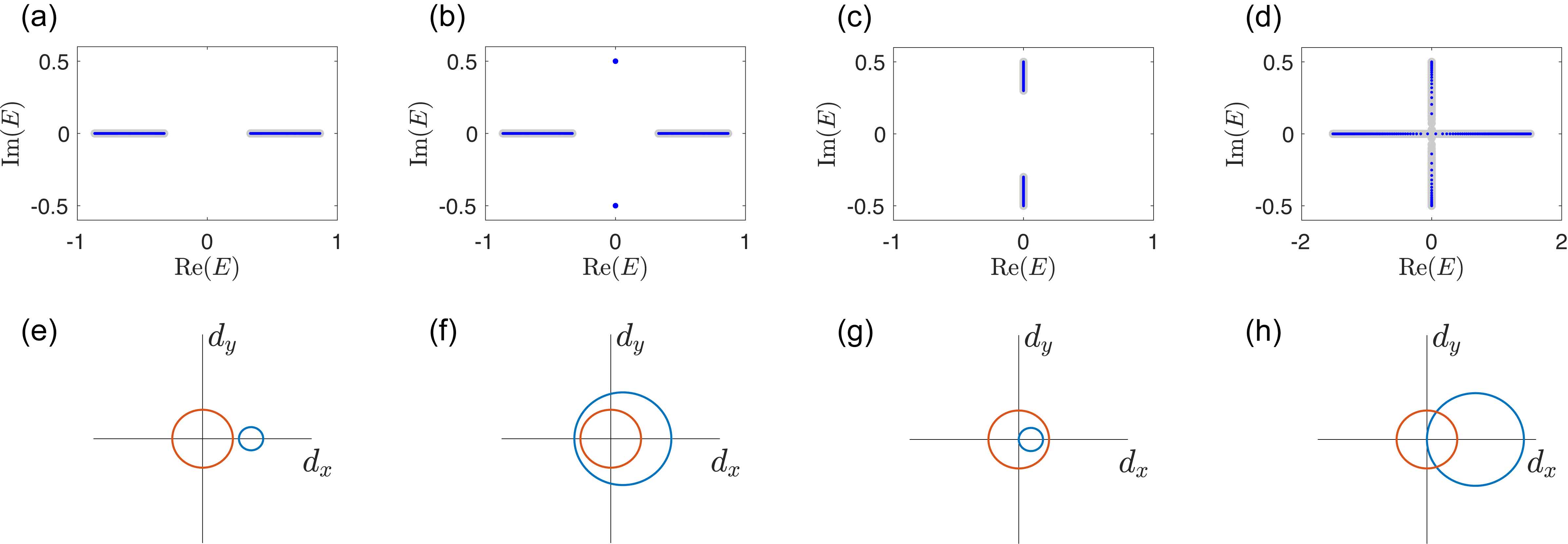}
\caption{Spectra and SR configurations of the {C-symmetric SSH model}. In each panel from (a) to (d), gray and dark blue colors indicate the PBC and OBC spectra respectively. (e)-(h) The geometric features corresponding to the spectra in (a)-(d) respectively, the red loops are the SRs, and the light blue loops are the ${\bm d}(k)$-loops. The parameters are $g_z=0.5$, with (a) $t_1=0.8$, $t_2=0.2$; (b) $t_1=0.2$, $t_2=0.8$; (c) $t_1=t_2=0.2$; and (d) $t_1=t_2=0.8$, \CH{corresponding to topologically trivial, nontrivial, non-quantized and gapless cases. In (b), the number of topological modes correspond to the $Z$ winding number of the ${\bm d}(k)$-loops around the SR loop.}}
\label{spectrum_SSH_type1}
\end{figure*}

For a two-component Hamiltonian $H(k)=({\bm d}+i{\bm g})\cdot{\bm \sigma}$, the vectors ${\bm d}$ and ${\bm g}$ are orthogonal to each other in the presence of CS, i.e. $\Gamma^{-1}H^{\dagger}(k)\Gamma=-H(k)$, \blue{as a complex term $(d_n+ig_n)\sigma_n$ cannot be mapped to its complex conjugation $(d_n-ig_n)\sigma_n$ with a unitary operator $\Gamma$.}
As a consequence, the ${\bm d(k)}$-loop and the SR (with its normal vector given by ${\bm g}$) lie in the same plane. \CH{The system is thus free from the NHSE, and shall exhibit the conventional BBC. As an example,} we consider a {``C-symmetric SSH model"} described by
\begin{eqnarray}
{\bm d}&=&(t_1+t_2\cos{k},t_2\sin{k},0)\nonumber\\
{\bm g}&=&(0,0,g_z),\label{SSH_type1}
\end{eqnarray}
with CS operator given by $\Gamma=\sigma_z$. In this model, the ${\bm d(k)}$-loop and the SR are both in the $d_x$-$d_y$ plane, and they can fall into four topologically different configurations, as shown in Fig.~\ref{spectrum_SSH_type1} with their typical spectra.
Figs.~\ref{spectrum_SSH_type1}(a) and (b) \CH{illustrate topologically trivial and nontrivial scenarios respectively, and are analogous to Hermitian SSH cases since the SR can be continuously shrunk towards} the origin without touching the ${\bm d(k)}$-loop. The topologically nontrivial case of (b) has a pair of edge states with opposite imaginary eigenenergies. In Fig.~\ref{spectrum_SSH_type1}(c), the ${\bm d(k)}$-loop is enclosed by the SR, corresponding to a purely imaginary spectrum.
\CH{Due to their geometrically distinct SR configurations, these three gapped phases cannot in general be directly transformed into each other, unless they go through a gapless phase like} in Fig.~\ref{spectrum_SSH_type1}(d). In this phase, the system has pairs of EPs protected by CS, which can only be gapped out when they merge into each other and annihilate.

\begin{figure}
\includegraphics[width=1\linewidth]{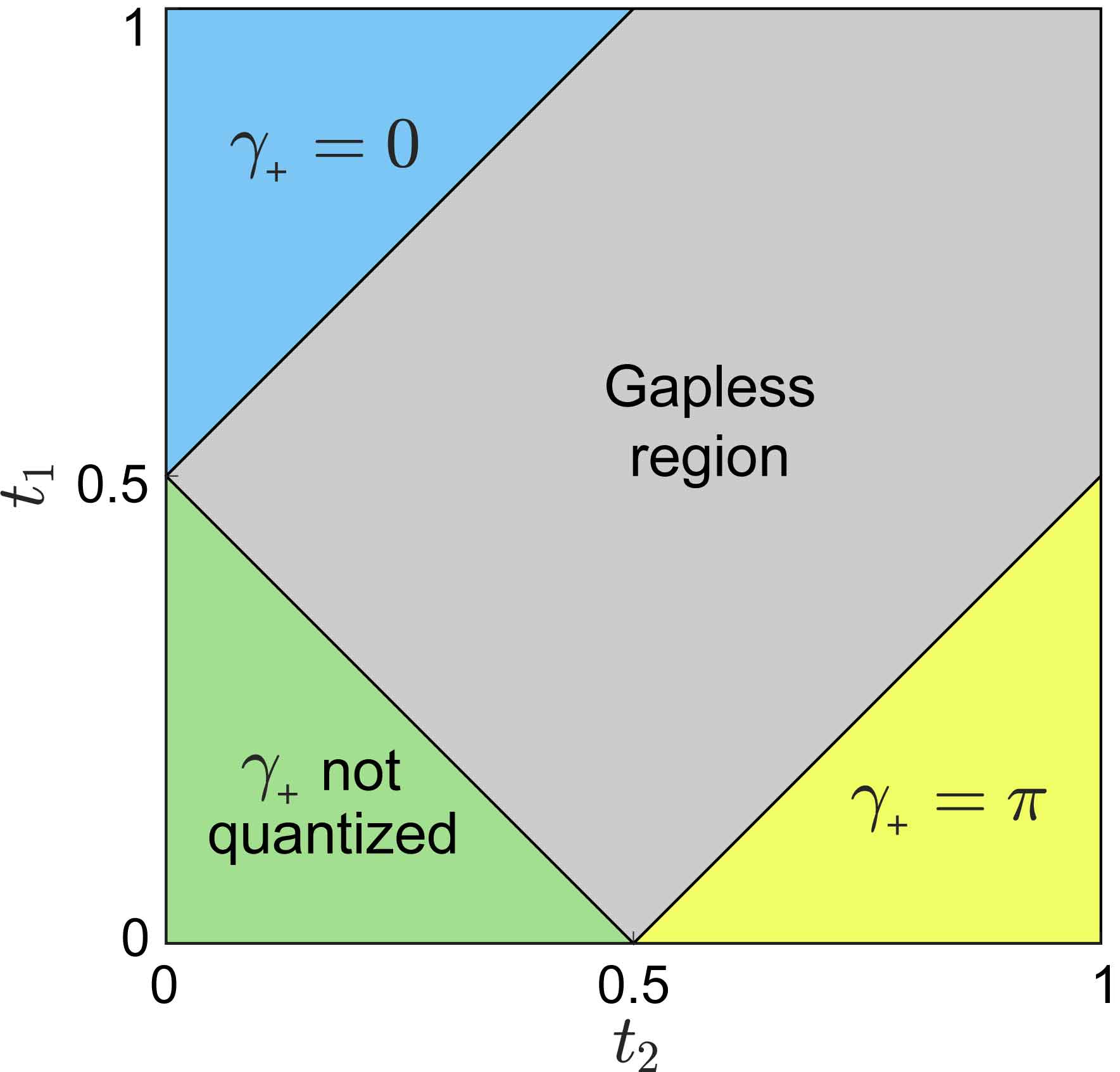}
\caption{Phase diagram of the single-band Berry phase [Eq.~(\ref{berryphase})] of the C-symmetric  SSH model with \CH{$g_z=0.5$}.}
\label{phase_SSH_type1}
\end{figure}

Since the BBC is preserved, we can calculate the Berry phase $\gamma_\pm$ as defined in Eq. (\ref{Berry_phase}) for a single band to distinguish the above configurations. \CH{Although their sum $\gamma_{\rm sum}$ is trivial since both the SR and the $\bold d(k)$ loop are confined to the same plane, individually they can still distinguish between the various possible phases.}
With some straightforward calculations, the Berry phase can be expressed as
\begin{eqnarray}
\gamma_\pm=-{\rm Im}\oint_0^{2\pi}dk \left[-i\frac{1\pm\cos \varphi}{2}\frac{\partial \theta}{\partial k}\right],
\label{berryphase}
\end{eqnarray}
with $\cos \varphi=ig_z/E_+$, $\cos \theta =d_x/\sqrt{d_x^2+d_y^2}$ \CH{and} $E_+=\sqrt{d_x^2+d_y^2-g_z^2}$.
In the case with a real $E_+$, the Berry phase is solely given by the winding of the ${\bm d}$ vector in the $d_x-d_y$ plane, and unaffected by the non-Hermitian term $g_z$. However, in the case with an imaginary $E_+$, since now $\cos \varphi$ takes a real value, $g_z$ acts like a real mass term and contributes a $k$-dependent coefficient to the derivation of $\theta$, leading to a non-quantized Berry phase.
Consistently, for each case of Fig.~\ref{spectrum_SSH_type1}, we find that $\gamma_\pm$ takes zero in (a), $\pi$ in (b), a non-quantized value in (c), and is ill-defined in (d) as the two bands become gapless.
A phase diagram depicting the Berry phase ${\gamma_+}$ is shown in Fig.~\ref{phase_SSH_type1}.
\blue{For the insulating phases with a real spectrum, the Berry phase can be expressed as $\gamma_{\pm}=\pi\nu_{\rm CS}$, with
\begin{eqnarray}
\nu_{\rm CS}=\frac{1}{2\pi}\oint_0^{2\pi}dk \frac{\partial \theta}{\partial k}
\end{eqnarray}
the winding number corresponding to how many times that the ${\bm d(k)}$-loops encloses SR, leading to a $Z$-type topological classification of systems with the same symmetry.}

\subsection{Sublattice symmetry (SLS)}\label{sec:SLS}
\CH{The SLS is characterized} by $\mathcal{S}^{-1}H(k)\mathcal{S}=-H(k)$, \blue{it involves no complex conjugation, hence if a term of $d_n\sigma_n$ can be mapped to $-d_n\sigma_n$ with a unitary operator $\mathcal{S}$, so can $ig_n\sigma_n$.}
This ensures that the vectors ${\bm d}$ and ${\bm g}$ lie in the same plane for each $k$. Therefore the SR is always perpendicular to the plane containing the ${\bm d(k)}$-loop, and \CH{intersect} the plane at two ``singularity points".

Due to symmetry protection, a winding number can be defined for each of these two singularity points, namely as how many times the ${\bm d(k)}$-loop winds around it~\cite{Yin2018nonHermitian}.
The difference between these two winding numbers give the vorticity and indicates the linking number of the ${\bm d(k)}$-loop and the SR. On the other hand, the sum of them is closely related to the number of edge states of a semi-infinite system~\cite{Yin2018nonHermitian}.
However, SLS does not eliminate the NHSE, suggesting that the winding numbers and the edge states under OBCs may not always match each other.
\CH{In particular}, the NHSE is unavoidable in the case with a nonzero linking number, as the complex spectrum must form some loops enclosing the origin on the complex plane.

To give a concrete example, we consider a {``SL-symmetric non-Hermitian SSH model"} described by
\begin{eqnarray}
{\bm d}&=&(t_1+t_2\cos{k},t_2\sin{k},0)\nonumber\\
{\bm g}&=&(g_x,g_y,0),
\end{eqnarray}
which satisfies SLS with $\mathcal{S}=\sigma_z$. In this model, $g_y$ corresponds to a non-Hermitian non-reciprocal hopping term, which induces the NHSE~\cite{Yao2018nonH1D,Martinez2018nonH,Lee2018Anatomy}. To \CH{uncover the geometric origin of the} topological edge states in this model, we need to consider the \CH{non-Bloch} Hamiltonian $\bar{H}(k)=H(k+i\kappa)$, which recovers the OBC \CH{skin} spectrum and is free of the NHSE~\cite{Yao2018nonH1D,Lee2018Anatomy}.
\blue{With some further analysis in Appendix \ref{app:SLS}, we obtain that the spectrum forms a doubly degenerate arc when
\begin{eqnarray}
e^{4\kappa}=\frac{t_1^2+g_y^2+2t_1g_y+g_x^2}{t_1^2+g_y^2-2t_1g_y+g_x^2},
\end{eqnarray}
which determines $\kappa$ of the {SL-symmetric SSH model.}}

\begin{figure}
\includegraphics[width=1\linewidth]{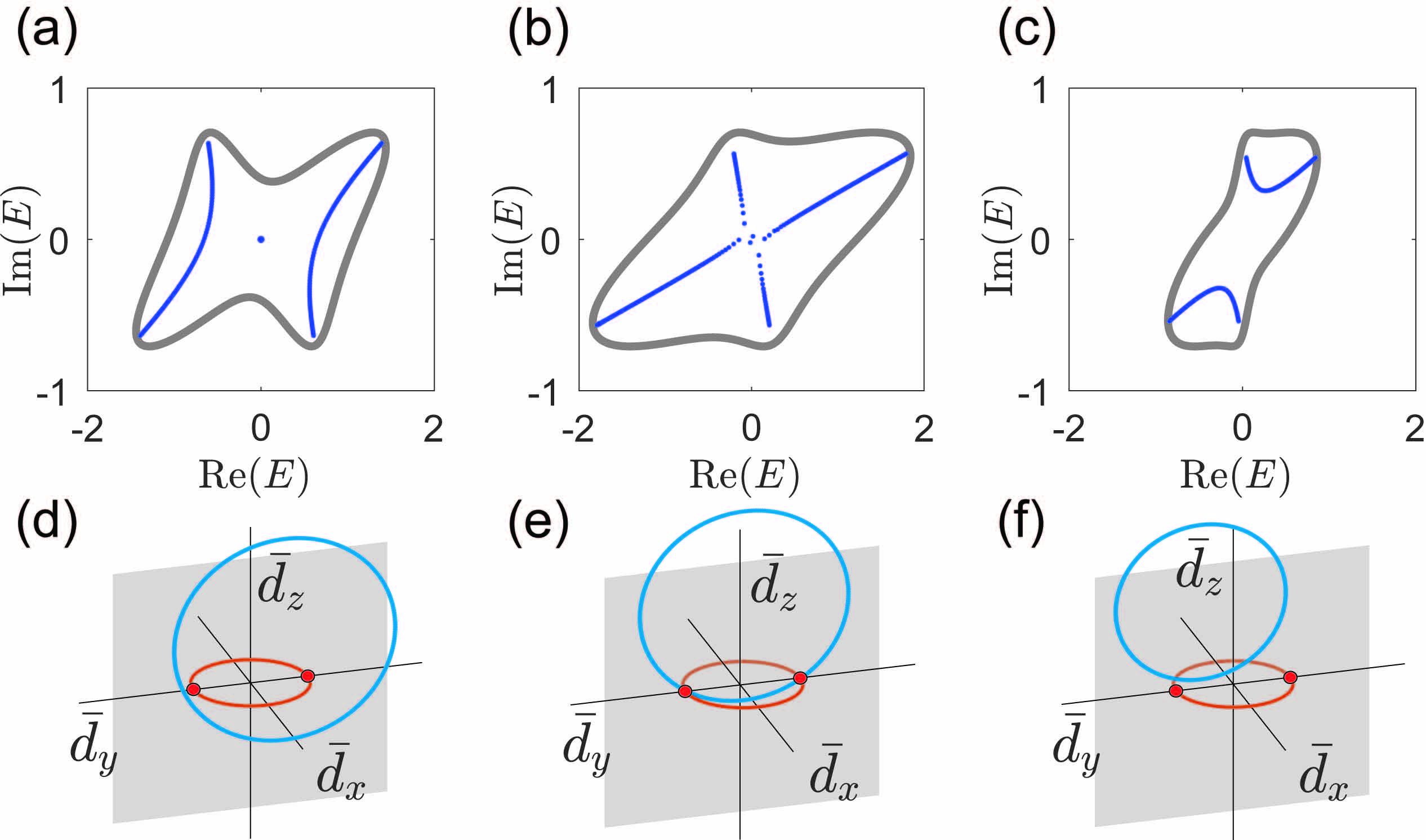}
\caption{Spectrum and SR winding pictures of the {SL-symmetric SSH model}, \CH{with (a-c) respectively corresponding to topologically trivial, gapless and nontrivial}. In each panel, gray and dark blue curves indicate the PBC and OBC spectra respectively. (d-f) the geometric features corresponding to (a-c), the red loops are the SR, and \CH{in light blue are the $\bar{\bm d}(k)$-loops which are confined to the gray planes.} The red dots are the singularity points in those planes.
Parameter used are $g_x=g_y=0.5$ and $t_2=1$, with (a) $t_1=0.5$; (b) $t_1=0.9$; (c) $t_1=1.2$, \CH{which correspond to topological, gapless and trivial phases respectively.} }
\label{spectrum_SSH_type2}
\end{figure}

In Fig.~\ref{spectrum_SSH_type2}, we illustrate the PBC and OBC spectra for the original $H(k)$, and the geometric features of its corresponding $\bar{H}(k)$. \CH{Note that the PBC and OBC spectra of $\bar{H}(k)$ coincide with the OBC spectrum of $H(k)$, with the exception of the topological modes.
This system has two topologically distinct} phases, with the $\bar{\bm d}$-loop enclosing either both or none of the two singularity points, as shown in Fig.~\ref{spectrum_SSH_type2}(a) and (c) respectively.
A phase transition between the two phases can be determined by requiring $\bar{E}_{\pm}=0$, which yields
\begin{eqnarray}
\cos{(k-\alpha)}=-f_0\sqrt{f_1^2+f_2^2}=-2t_1g_x\sqrt{f_3^2+f_4^2}.\qquad
\end{eqnarray}
At this transition point, the ${\bm d(k)}$-loop shall touch the SR at two points, as in Fig.~\ref{spectrum_SSH_type2}(e).

{Note that since the non-Hermitian term $\bar{\bm g}$ in $\bar{H}(k)$ depends on $k$, the $\bar{\bm d}$-loop and the SR have to be obtained with the normalization procedure introduced in Section~\ref{sec:normalization}. As such}, in Fig.~\ref{spectrum_SSH_type2}(d-f), the $\bar{\bm d}$-loop and the SR do not have to satisfy the conditions of Eqs. (\ref{no_skin_con1}) and (\ref{no_skin_con2}), even though the NHSE has been removed by the complex deformation $k\rightarrow k+i\kappa$. \CH{This is because the normalization distorts the geometry while still keeping the topological information.}

For a more general model with SLS, the $\bar{\bm d}(k)$-loop can wind around the two singularity points multiple times, \CH{giving rise to a} $Z$-type topological classification of the system.
It is worth mentioning that while the ${\bm d(k)}$-loop of the original Hamiltonian $H(k)$ can link with the SR and enclose the two singularity points a different number of times, the $\bar{\bm d}$-loop cannot as the effective Hamiltonian $\bar{H}(k)$ does not possess NHSE, and cannot have nontrivial vorticity.

\subsection{Conjugated particle-hole symmetry (PHS$^\dagger$)}\label{sec:PHSdag}
Having discussed the two non-Hermitian extensions of Hermitian $Z$-type topology protected by chiral symmetry,
{we next turn to the $Z_2$ topology protected by PHS and its non-Hermitian variant PHS$^\dagger$, both of which have received less attention in the study of non-Hermiticity.} In Hermitian system, PHS ensures that the pseudospin vector always winds a solid angle of either $0$ or $2\pi$,
corresponding to a Berry phase of $0$ or $\pi$~\cite{Li2016EPJB}.
Here, we first consider the variant PHS$^{\dagger}$ described by $\mathcal{T}_-^{-1}H^*(k)\mathcal{T}_-=-H(-k)$, \CH{which also restricts the behavior of the ${\bm d(k)}$-loop. For illustration, we consider an example given by a \blue{``PH$^\dagger$-symmetric''} non-Hermitian SSH model" }
\begin{eqnarray}
{\bm d}&=&(t_1+t_2\cos{k},t_2\sin{k},0)\nonumber\\
{\bm g}&=&(0,g_y,g_z),
\label{typeIII}
\end{eqnarray}
with the CS and the SLS previously discussed broken by $g_y$ and $g_z$ respectively. \CH{Nevertheless, because the Hermitian part ${\bm d}\cdot {\bm \sigma}$ belongs to the Hermitian BDI symmetry class and already satisfies a PHS, we shall name the resultant symmetry of this non-Hermitian model as PHS$^{\dagger}$,} with $\mathcal{T}_-=\sigma_z$. {While this model has already been investigated in detail in Ref.~\onlinecite{Hui2018nonH}, we shall unveil how it can be further characterized geometrically} with the SR.

\begin{figure}
\includegraphics[width=1\linewidth]{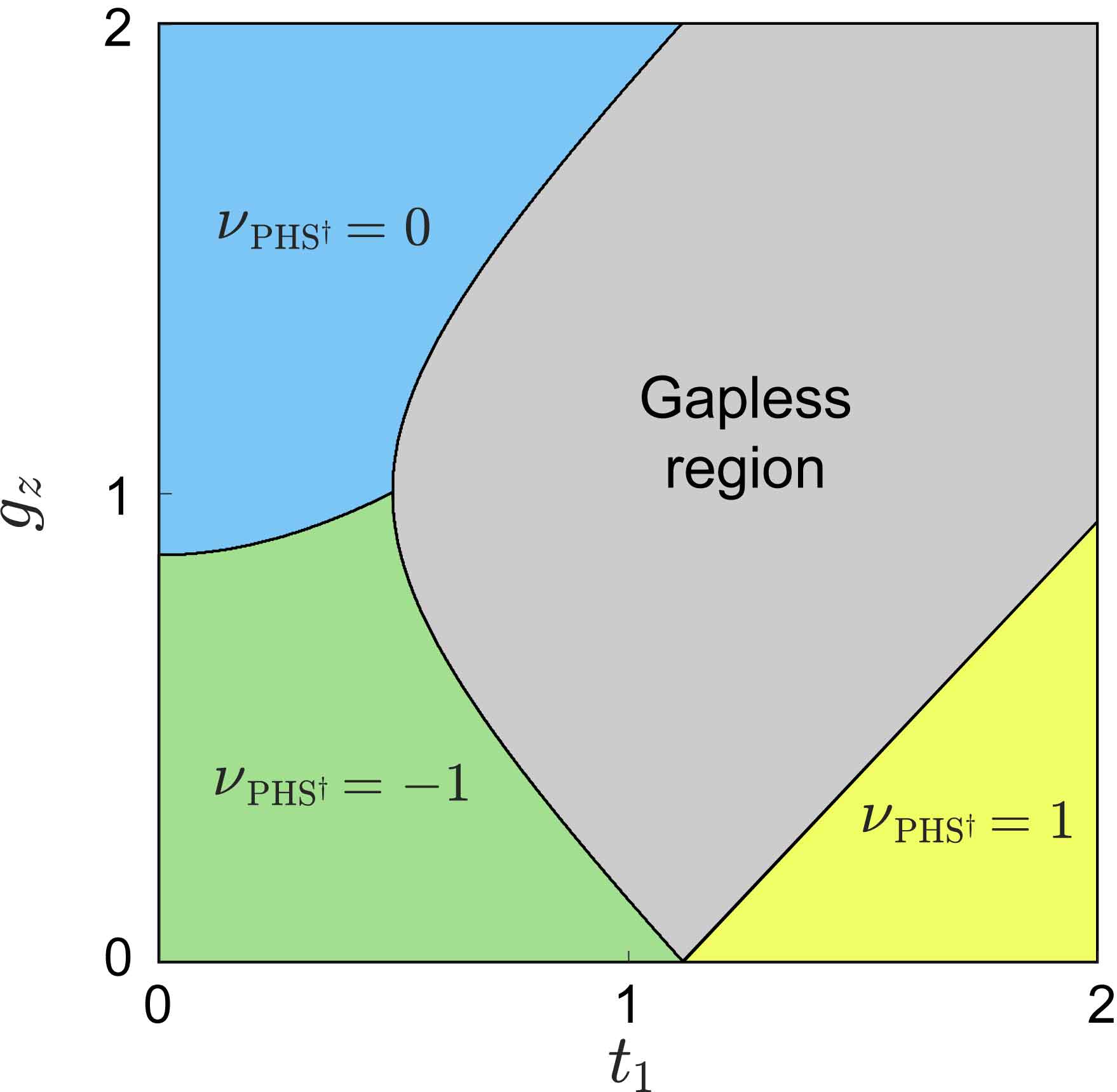}
\caption{Phase diagram of $\nu_{\text{PHS}^\dagger}$ [Eq.\ref{nu3}] for the \blue{PH$^\dagger$-symmetric SSH model} with fixed $\gamma_y=0.5$ and $t_2=1$. }
\label{phase_SSH_type3}
\end{figure}

\begin{figure*}
\includegraphics[width=1\linewidth]{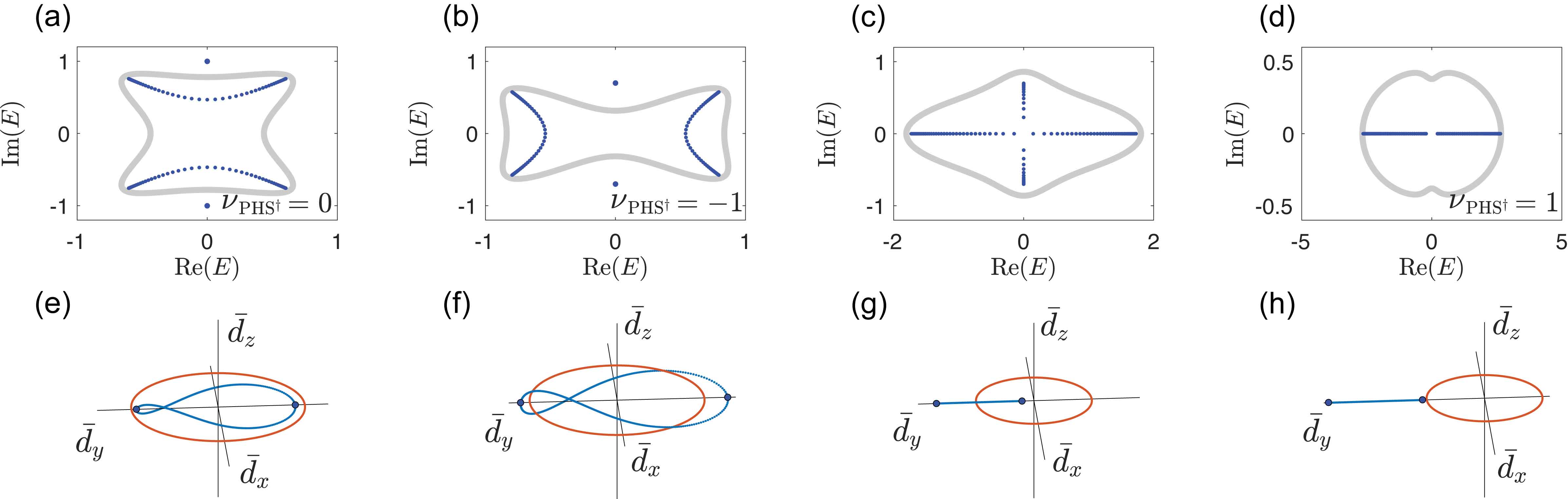}
\caption{Spectrum and SR winding behavior of the \blue{PH$^\dagger$-symmetric SSH model}. In (a-d), gray and dark blue curve indicate the PBC and OBC spectra respectively, \CH{with the PBC spectra bearing no direct influence on the presence and position of topological modes.}  In (e-h), the red loops are the SR, and in light blue ones are the $\bar{\bm d}$-loops, corresponding to (a-d) respectively. \blue{The blue dots indicate the high-symmetric points, which fall onto the $\bar{d}_y$-axis after normalizing the SR.
(e) $\nu_{\text{PHS}^\dagger}=0$ when the two high-symmetric points are inside the SR.
(f,h) When the two high-symmetric points are outside the SR, $\nu_{\text{PHS}^\dagger}$ takes $-1$ or $1$, depending on whether the two points are separated by the SR along $\bar{d}_y$-axis or not.}
The parameters are \CH{$g_y=0.5$} and $t_2=1$, with (a) $t_1=0.2$, $g_z=1$; (b) $t_1=0.2$, $g_z=0.7$; (c) $t_1=1$, $g_z=0.7$; and (d) $t_1=1.8$, $t_2=0.7$. }
\label{spectrum_SSH_type3}
\end{figure*}

Due to the non-reciprocal coupling from $g_y$, Eq.~\ref{typeIII} also possesses NHSE. Hence we consider the \CH{non-Bloch} Hamiltonian with $k\rightarrow k+i\kappa$, described by
\begin{eqnarray}
\bar{H}(k)=(\bar{\bm d}+i\bar{\bm g})\cdot{\bm \sigma},
\end{eqnarray}
which gives a doubly degenerate spectrum when
\begin{eqnarray}
e^{4\kappa}=\frac{t_1^2+g_y^2+2t_1g_y}{t_1^2+g_y^2-2t_1g_y},\label{kappa_SSH3}
\end{eqnarray}
with detailed derivation given in Appendix \ref{app:PHSdag}.
Furthermore, the explicit form of the non-Bloch Hamiltonian in Appendix \ref{app:PHSdag}
indicates that the $\bar{\bm d}$-loop has a two-fold rotation symmetry regarding $\bar{d}_x$-axis, and both it and the SR must cross the $\bar{d}_x$-axis at the two high-symmetric points of $k_0=0$ or $\pi$.
Also, the $\bar{\bm d}$-loop cannot link to the SR since $\bar{H}(k)$ does not have NHSE. Due to these results, there are only three different gapped phases that cannot transform into each other without closing the band gap: at both the two high-symmetric points, the non-vanishing terms satisfy
\begin{eqnarray}
i)&~~~~&|\bar{d}_x(k_0)|>r_{\bar{g}}(k_0)~{\rm and}~\bar{d}_x(0)\bar{d}_x(\pi)>0; \nonumber\\
ii)&~~~~&|\bar{d}_x(k_0)|<r_{\bar{g}}(k_0);~{\rm and}\nonumber\\
iii)&~~~~&|\bar{d}_x(k_0)|>r_{\bar{g}}(k_0)~{\rm and}~\bar{d}_x(0)\bar{d}_x(\pi)<0, \nonumber
\end{eqnarray}
with $r_{\bar{g}}(k)=\sqrt{\bar{g}^2_y(k)+\bar{g}^2_z(k)}$ the radius of the normalized SR.
These three cases can be distinguished by a topological invariant defined as
\begin{eqnarray}
\nu_{\text{PHS}^\dagger}={\rm Sign}\{{\rm Re}[\bar{S}_x(0)][{\rm Re}[\bar{S}_x(\pi)]\}
\label{nu3}
\end{eqnarray}
which takes $1$, $0$, and $-1$ respectively, with $\bar{S}_x(k)=\bar{d}_x(k)/\bar{E}_+(k)$, \blue{$\bar{E}_+(k)$ a non-Bloch eigen-energy which is imaginary in case (ii), and real in cases (i) and (iii).} This number $\nu_{\text{PHS}^\dagger}$ provides a $Z_2$ classification of the two topological insulating phases.
In Fig.~\ref{phase_SSH_type3} we illustrate the phase diagram of this model, which also contains a gapless region where all the eigenenergies are either purely imaginary or real. In such cases, we find that the two high-symmetric points have imaginary and real eigenenergies respectively.

\blue{In Fig.~\ref{spectrum_SSH_type3}, we illustrate the spectra and the geometrical properties of this model.
As now $\bar{\bm g}$ has a dependence of $k$, we need to consider the normalization of SR in Sec. \ref{sec:normalization}, which also rotates the high-symmetric points from $\bar{d}_x$-axis to $\bar{d}_y$-axis.}
In Fig.~\ref{spectrum_SSH_type3}(b) with $\nu_{\text{PHS}^\dagger}=-1$, there is a pair of edge states protected by an imaginary line-gap. These edge states may not disappear when the two bands touch and become gapped by a real line-gap [with $\nu_{\text{PHS}^\dagger}=0$ as in Fig.~\ref{spectrum_SSH_type3}(a)], but it is no longer protected by the gap. In all the four cases, the PBC spectra assume qualitatively similar closed loops, but their OBC spectra are topologically distinct, and can be well characterized by the invariant $\nu_{\text{PHS}^\dagger}$.

\blue{Finally, it is also observed that the spectra in Fig. \ref{spectrum_SSH_type3}(c,d) fall only onto the real and imaginary axis, as the normalized $\bar{\bm d}$-loop becomes a line on the $\bar{d}_x$ axis in the vector space.
To understand this, we note that the non-Bloch Hamiltonian in Appendix \ref{app:PHSdag} satisfies $\bar{\bm d}(k)\cdot\bar{\bm g}(k)=0$ for any $k$ when $t_1>g_y>-t_1$ (or when $t_1<g_y<-t_1$), hence it recovers an effective CS and its spectrum behaves similarly to the C-symmetric SSH model discussed in Section \ref{sec:CS}.}


\subsection{Particle-hole symmetry (PHS)}\label{sec:PHS}
\begin{figure}
\includegraphics[width=1\linewidth]{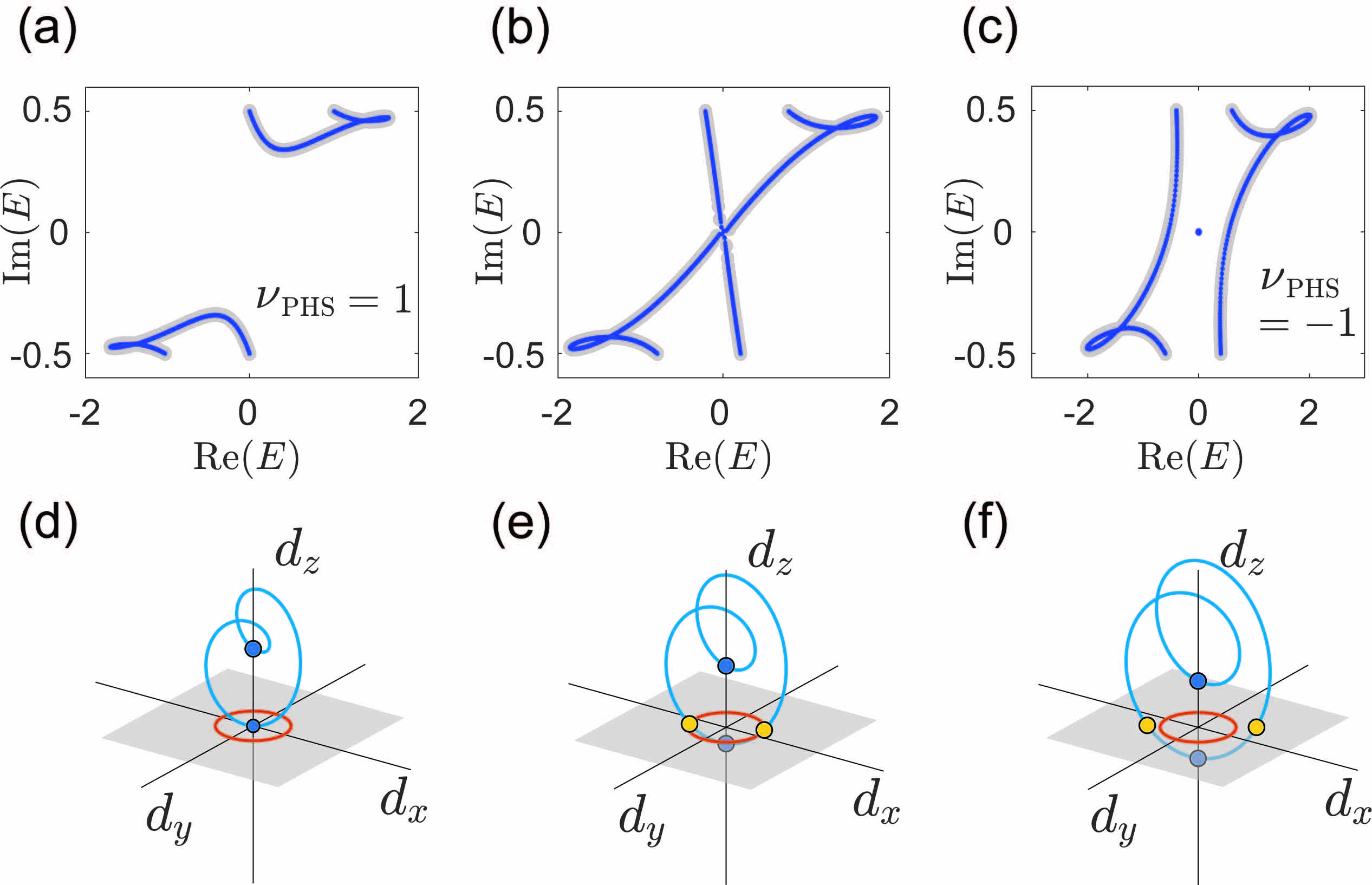}
\caption{Spectrum and SR winding visualizations of the \CH{PH-symmetric} non-Hermitian \CH{extended} Kitaev model, \CH{with cases (a-c) being topologically trivial, gapless and nontrivial respectively}. In (a-c), gray and dark blue curves indicate the PBC and OBC spectra respectively. In (d-f), the red loops are the SR, and in light blue are the ${\bm d(k)}$-loops, corresponding to (a-c) respectively. The gray areas in (d-f) are the planes containing the SR. The blue dots correspond to the high-symmetric points, \blue{and the yellow ones indicate where the ${\bm d(k)}$-loops penetrate the SR-plane.}
The parameters are $\gamma_z=0.5$, $\phi=\pi/2$, $\mu=1$, $t_1=\Delta_1=0.5$, with (a) $t_2=\Delta_2=0.5$; (b) $t_2=\Delta_2=0.711$; (c) $t_2=\Delta_2=0.9$.}
\label{spectrum_Kitaev_type1}
\end{figure}

PHS is described by $\mathcal{C}_-^{-1}H^T(k)\mathcal{C}_-=-H(-k)$. With some straightforward analysis, we can see that it is impossible to preserve the PHS while breaking both the CS and SLS with non-Hermitian terms of $(g_x,g_y,g_z)$ in the SSH model. In \CH{the non-Hermitian setting, PHS enforces}
\begin{eqnarray}
d_x(k)+ig_x(k)&=&-d_x(-k)-ig_x(-k),\nonumber\\
d_y(k)+ig_y(k)&=&-d_y(-k)-ig_y(-k),\nonumber\\
d_z(k)+ig_z(k)&=&d_z(-k)+ig_z(-k),
\end{eqnarray}
which yields $E_\pm(k)=E_\pm(-k)$, ensuring that the system is always free from NHSE. Without loss of generality, we consider $k$-independent non-Hermitian terms as before, which yields $g_x=g_y=0$ as required by PHS. Therefore, \CH{for an illustrative model}, we consider a non-Hermitian extension of the Kitaev model~\cite{Kitaev,Li2016EPJB} given by
\begin{eqnarray}
d_x&=&\Delta_2\sin\phi\sin 2k,\nonumber\\
d_y&=&\Delta_1\sin k+\Delta_2\cos\phi\sin 2k,\nonumber\\
d_z&=&-t_1\cos k-t_2\cos 2k+\mu,\nonumber\\
{\bm g}&=&(0,0,g_z),
\end{eqnarray}
with the PHS \CH{operator} given by $\mathcal{C}_-=\sigma_x$. The absence of chiral symmetry is ensured by the next-nearest-neighbor couplings $t_2$ and $\Delta_2$ with a non-zero phase $\phi$, leaving the zero edge states solely protected by PHS~\cite{Li2016EPJB}. Restricted by PHS, the ${\bm d(k)}$-loop must be two-fold rotationally symmetric \CH{about the $d_z$-axis, and the SR must lie on the $d_x$-$d_y$ plane.} Thus the system only has two topologically different phases, depending on whether the ${\bm d(k)}$-loop ``encloses" the SR or not, as shown in Fig.~\ref{spectrum_Kitaev_type1}(a) and (c). More specifically, the system has a pair of zero edge modes when the ${\bm d(k)}$-loop intersects the $d_x$-$d_y$ plane outside the SR, and has no edge modes otherwise (intersects inside the SR, or does not touch the plane).
These two cases can be \CH{regarded as deformations of the two topologically distinct phases of the Hermitian extended Kitaev model,
whose topology can be characterized by a $Z_2$ quantity defined by the two high-symmetry} points $k=0$ and $\pi$ as
\blue{\begin{eqnarray}
\nu_{\rm H-PHS}={\rm Sign}[d_z(0)d_z(\pi)].
\end{eqnarray}}
However, in a non-Hermitian system,
the ${\bm d(k)}$-loop may touch the SR at any pair of $\pm k$, which induces a topological phase transition [Fig.~\ref{spectrum_Kitaev_type1}(b)].
Hence, in contrast to Hermitian systems with PHS, the two topologically different phases here cannot be distinguished solely by looking at the two high-symmetric points.

\blue{To characterize the topological properties of the system, we note that in Hermitian system, the quantity $\nu_{\rm H-PHS}$ can also be expressed as $(-1)^N$, with $N$ the number of times that the vector ${\bm h}(k)$ penetrates the $h_x-h_y$ plane when $k$ varies from $0$ to $\pi$. In our system, a nonzero $g_z$ induce the SR, but does not change the topological property until the SR touches the ${\bm d}$-loop. Therefore the topological invariant related to PHS can be simply defined as
\begin{eqnarray}
\nu_{\rm PHS}=(-1)^{N_+},
\end{eqnarray}
with $N_+$ the number of times that the vector ${\bm h}(k)$ penetrates the $d_x-d_y$ plane {\it outside the SR} when $k$ varies from $0$ to $\pi$.}

\section{2D extension}\label{sec:2D}

\begin{figure*}
\includegraphics[width=1\linewidth]{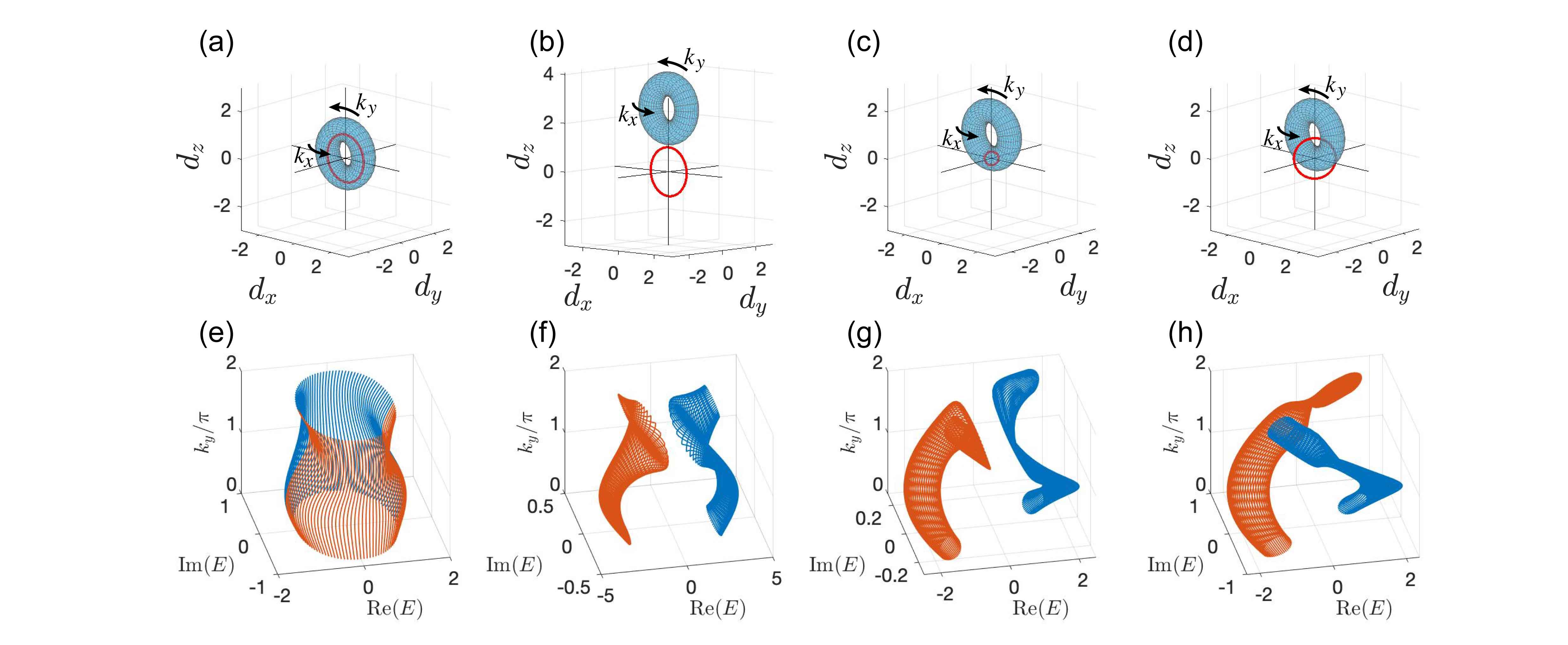}
\caption{\CH{Various distinct geometric configurations (a-d) and corresponding PBC spectra versus $k_y$ (e-h)} of the 2D model given by Eqs. (\ref{h2D},\ref{h2D_d}). The blue tori and the red loops in (a)-(d) represent the ${\bm d}$-torus and SR respectively. \CH{Cases (a,d) exhibit vorticities in the $k_x$ and $k_y$ directions respectively, but not cases (b,c) which are however topologically distinct.}
Parameters used are (a) $\mu=0.2$, $g_y=1$, $g_x=g_z=0$; (b) $\mu=2.6$, $g_y=1$, $g_x=g_z=0$; (c) $\mu=1$, $g_x=-g_y=0.2$, $g_z=0$; and (d) $\mu=1$, $g_x=-g_y=0.6$, $g_z=0$. Other parameters are $t_1=1$ and $t_2=0.5$.}
\label{spectrum_2D_untouched}
\end{figure*}
\subsection{Geometric features and PBC spectrum}
\CH{To further illustrate the usefulness of our SR geometric picture,} we extend our analysis of 1D systems to 2D cases, \CH{where geometric visualization provides greater insights}.
In 2D two-component systems, the \CH{trajectory of ${\bm d}(\bm k)$-vector throughout the Brillouin zone forms a 2D torus in a 3D vector space. Other than having robust band touchings with the SR, this 2D torus can also be in topologically distinct geometric configurations with the SR, resulting in qualitatively different band structure.}

To give an example, we consider a 2D system described by
\begin{eqnarray}
H_{\rm 2D}(k_x,k_y)=\sum_{n=x,y,z}(d_n(k_x,k_y)+ig_n)\sigma_n,\label{h2D}
\end{eqnarray}
with
\begin{eqnarray}
d_x(k_x,k_y)&=&(t_1+t_2\cos{k_x})\cos{k_y}\nonumber\\
d_y(k_x,k_y)&=&t_2\sin{k_x}\nonumber\\
d_z(k_x,k_y)&=&(t_1+t_2\cos{k_x})\sin{k_y}+\mu.
\label{h2D_d}
\end{eqnarray}
The winding of ${\bm d}(k_x,k_y)$ vector forms a prefect torus, with the major and minor radii given by $t_1$ and $t_2$ respectively, and the center of the torus given by $\mu$.
Other than touching each other, there are four \CH{distinct geometric possibilities} between the ${\bm d}$-torus and the SR under different parameters, as shown in Fig.~\ref{spectrum_2D_untouched}. In Fig.~\ref{spectrum_2D_untouched}(a), the SR runs through the tube of the ${\bm d}$-torus. Here the ${\bm d}(k_x,k_y)$ vector winds around the SR when $k_x$ varies \CH{through} a period with fixed $k_y$, \CH{but not when $k_y$ is instead varied with fixed $k_x$.} Hence its corresponding spectrum versus $k_y$ forms a cylinder-like manifold as shown in Fig.~\ref{spectrum_2D_untouched}(e), indicating nontrivial linkage (half-integer vorticity) along $k_x$ but not $k_y$.
\CH{In Fig.~\ref{spectrum_2D_untouched}(b), the ${\bm d}$-torus is completely separated from the SR, while in Fig.~\ref{spectrum_2D_untouched}(c), it completely encloses the SR such that the latter can be shrunk into a point}. \CH{In both cases (b,c), any 1D \CH{closed} path on the torus cannot link with the SR, resulting in two PBC bands that are fully separated from each others (zero vorticity), as illustrated} in Fig.~\ref{spectrum_2D_untouched}(f,g).
On the other hand, the SR can also enclose the ${\bm d}$-torus [\ref{spectrum_2D_untouched}(d)], \CH{which gives the case of~\ref{spectrum_2D_untouched}(a) but with the roles of $k_x$ and $k_y$ exchanged}: the ${\bm d}(k_x,k_y)$ vector instead winds around the SR when $k_y$ varies through a period with fixed $k_x$, but not when $k_x$ varies with $k_y$ being fixed. Consequently, the spectrum versus $k_y$ gives two separate bands for each fixed $k_y$, but  \CH{which evolve into each other when $k_y$ varies over} a period of $2\pi$ [\ref{spectrum_2D_untouched}(h)].

\begin{figure}
\includegraphics[width=1\linewidth]{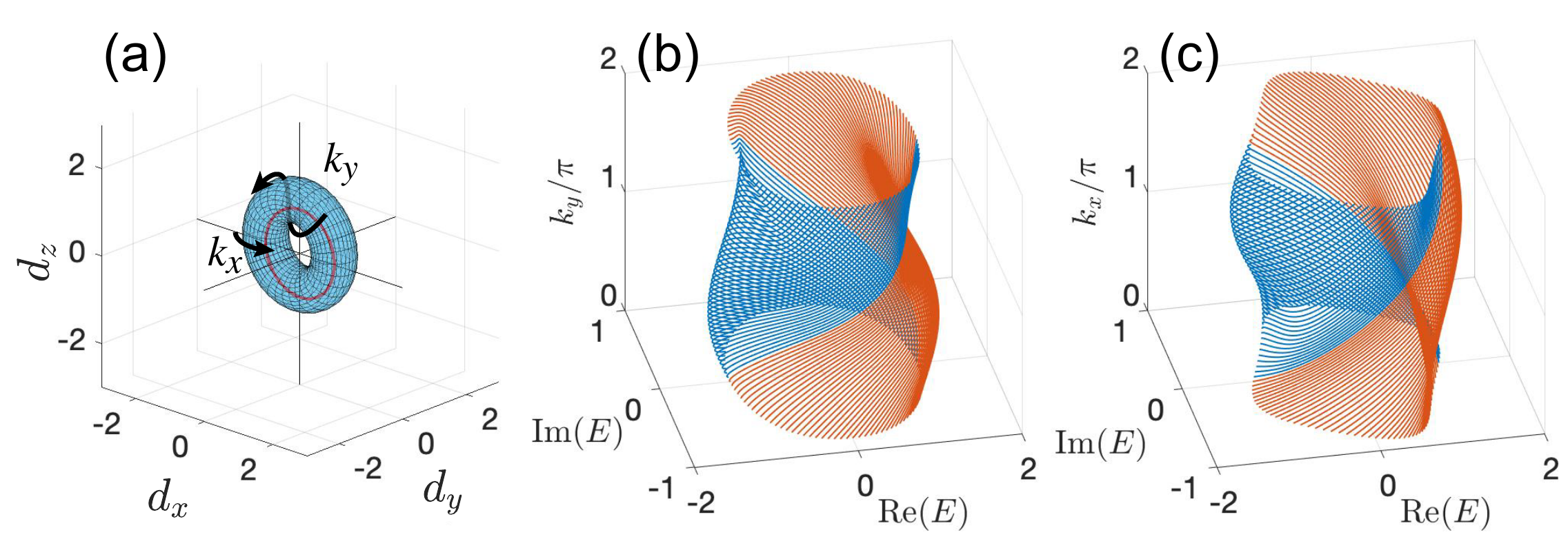}
\caption{Geometric configuration and spectra versus $k_y$ or $k_x$ of the 2D model given by Eq. (\ref{h2D_2}), \CH{where a Dehn twist around the torus gives rise to nontrivial vorticity in both directions}.
The parameters are $t_1=1$, $t_2=0.5$, $\mu=0.2$, $g_{x,z}=0$, $g_y=1$.}
\label{double_link}
\end{figure}

\CH{In the above, we have seen that the two bands can evolve into each other either along the $k_x$ or $k_y$ direction (half integer vorticity), or not evolve at all (integer vorticity).}
Naturally, one may ask whether the two bands can \CH{evolve into} each other along both $k_x$ and $k_y$ directions simultaneously.
Such a scenario is possible with a shift of the quasi-momentum, i.e. $k_x\rightarrow k_x+k_y$, of the Hamiltonian of Eqs.~\ref{h2D_d}.
\CH{In Fig.~\ref{double_link}, we illustrate
\begin{eqnarray}
H'_{2D}(k_x,k_y)=H_{2D}(k_x+k_y,k_y)\label{h2D_2}
\end{eqnarray}
for the case where the SR threads within the circumference of the ${\bm d}'(k_x,k_y)$ torus. 
Since ${\bm d}'(k_x,k_y)$ now winds around the SR along both the $k_x$ and $k_y$ directions,} the spectrum forms a cylinder-like manifold versus each quasi-momentum [Fig.~\ref{double_link}(b,c)]. \CH{In general, arbitrarily large vorticities can be obtained in either direction via a combination of generalized Dehn twists $k'_x=ak_x+bk_y$, $k'_y=ck_x+dk_y$ and intrinsic windings present in the mapping ${\bm d}(k'_x,k'_y)$, where $a,b,c,d$ are integers.} 

\subsection{Topological boundary states under OBCs}

 \begin{figure*}
\includegraphics[width=1\linewidth]{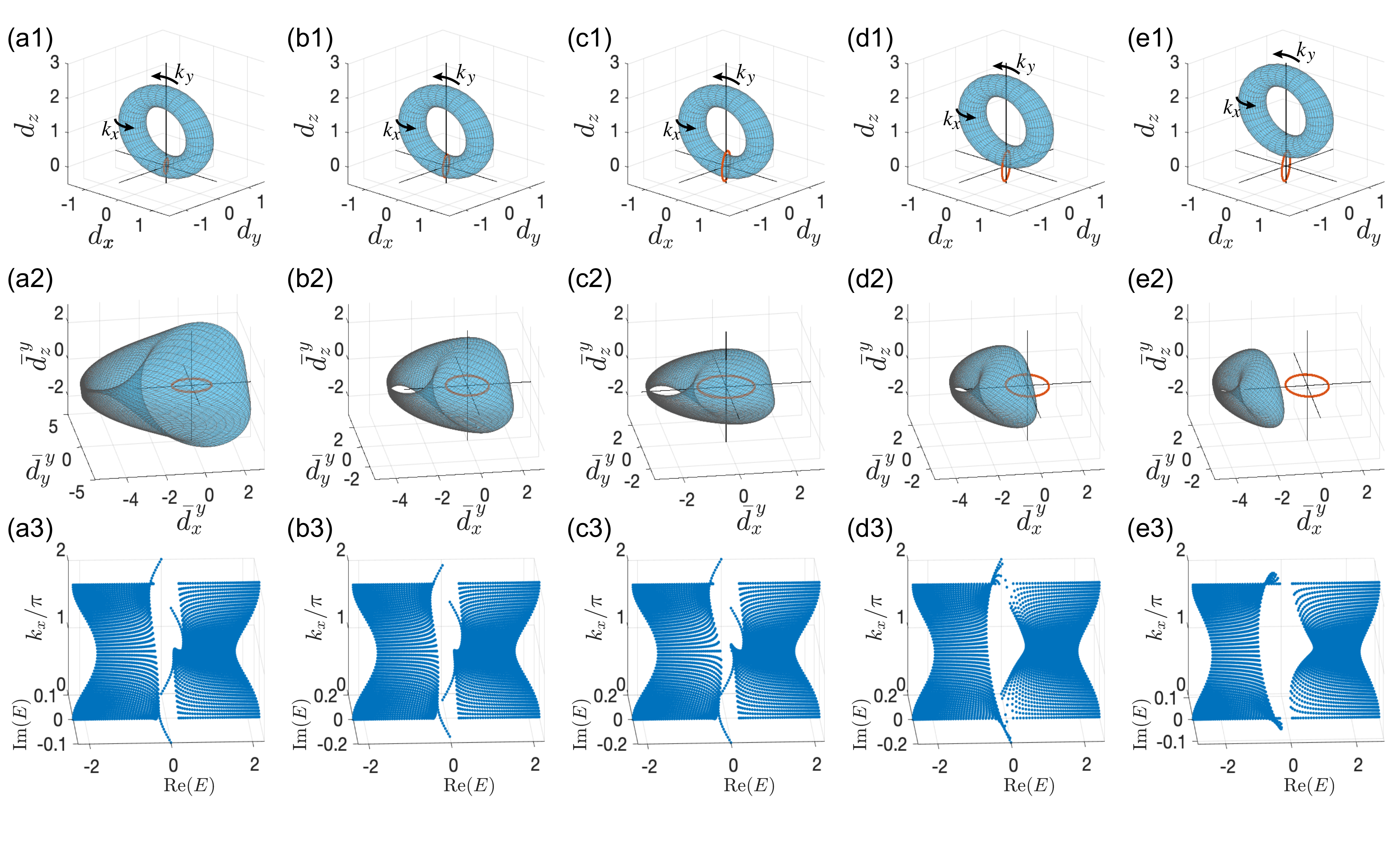}
\caption{Geometric visualization of the 2D effective model of $\bar{H}^y_{\rm 2D}(k_x,k_y)$, \CH{with distortion of the torus resulting from $k_y\rightarrow k_y+i\kappa_y$ and the consequently requisite normalization of the SR.} The parameters are (a) $g_0=0.2$, $\mu=1$, (b) $g_0=0.3$, $\mu=1$, (c) $g_0=0.4$, $\mu=1$, (d) $g_0=0.4$, $\mu=1.3$, (e) $g_0=0.4$, $\mu=1.6$, with $g_x=g_0\cos\frac{\pi}{6}$ and $g_y=g_0\sin\frac{\pi}{6}$. The other parameters are $t_1=1$, $t_2=0.3$, $g_z=0$.}
\label{geometry_ykx}
\end{figure*}

The two cases shown in Fig.~\ref{spectrum_2D_untouched}(b,c), while sharing similar PBC spectrum, are topologically distinct in their SR geometric visualizations. \CH{This distinction translates to topologically different OBC bands as well.} By continuously tuning the non-Hermtian terms to zero, the SR in these cases can shrink into \CH{a point at the origin without touching the ${\bm d}$-torus, where usual Bloch sphere homotopy arguments gives a Chern number of} $0$ and $\pm 1$ (or $-1$) respectively. 
A nonzero Chern number indicates chiral boundary states under OBCs, which are also inherited by the non-Hermitian system.
Similar to 1D systems, a topological phase transition marking the emergence or disappearance of chiral-like boundary states can occur only when the ${\bm d}$-torus and the SR are linked [Fig.~\ref{spectrum_2D_untouched}(a,d)], or when they intersect. In Fig.~\ref{geometry_ykx} we demonstrate a transition between cases with and without chiral-like boundary states with OBCs along $y$ direction, with the non-Hermitian terms set to
\begin{eqnarray}
g_x=g_0\cos(\pi/6),~g_y=g_0\sin(\pi/6),~g_z=0
\end{eqnarray}
for a clearer illustration. The SR is enclosed by the ${\bm d}$-torus \CH{such that it is homotopic to a} point in Fig.~\ref{geometry_ykx}(a1),
where the $y$-OBC spectrum in Fig.~\ref{geometry_ykx}(a3) has a pair of 1D boundary states connecting the two separated bands.
When $g_0$ increases, the SR will be enlarged and touch the ${\bm d}$-torus, and eventually link with it, as shown in Fig. \ref{geometry_ykx}(b1) and (c1). However, the $y$-OBC spectrum remains topologically unchanged as the two bands do not touch each other in this parameter region.
Next, we increase the value of $\mu$ for a better illustration of the transition of $y$-OBC boundary states. \CH{Doing so shifts the ${\bm d}$-torus along $d_z$, such that it intersects} with the SR in a certain parameter region [Fig.~\ref{geometry_ykx}(d1,e1)].
\blue{Touching of the two bands and hence a topological phase transition emerges when the SR intersects the ${\bm d}$-torus [Fig.~\ref{geometry_ykx}(d3)]. Further increasing $\mu$ results in a separated y-OBC bands without boundary states [Fig.~\ref{geometry_ykx}(e3)], while the geometric features remain the same in Fig.~\ref{geometry_ykx}(d1) and (e1). }

\CH{To relate the SR winding pictures to the OBC spectra, the NHSE must be taken into account by performing the analytic continuation $k_j\rightarrow k+i\kappa_j$, with $j$ being either $x$ or $y$, and $\kappa_j$ determined like in the 1D cases.}
We shall deal with $k_x$ and $k_y$ separately. We start from the PBC spectrum, and first consider the spectral flow of $\kappa_y$, with $\kappa_x$ set to zero. For the non-Bloch Hamiltonian
\begin{eqnarray}
\bar{H}^y_{\rm 2D}({\bm k})=H_{\rm 2D}({\bm k}+i\kappa_y),
\end{eqnarray}
we obtain [see Appendix \ref{app:2D_yOBC}]
\begin{eqnarray}
e^{4\kappa_y}=\frac{\mu^2+\gamma_x^2+\gamma_z^2-2\mu\gamma_x}{\mu^2+\gamma_x^2+\gamma_z^2+2\mu\gamma_x},
\end{eqnarray}
which \CH{does not depend on $k_x$}.
In Fig.~\ref{geometry_ykx}(a2-e2), we visualize geometric features of $\bar{H}^y_{\rm 2D}(k_x,k_y)$ throughout transition from Fig.~\ref{geometry_ykx}(a3) to (e3). \CH{The $\bar{d}^y({\bm k})$ torus looks distorted due to the combination of the $k_y\rightarrow k_y+i\kappa_y$ substitution and the necessary SR normalization that transfers the momentum dependence of the non-Hermitian terms to the Hermitian terms.} From Fig.~\ref{geometry_ykx}(c2) to (e2), the SR moves from the interior of the $\bar{\bm d}^y$-torus of $\bar{H}^y_{\rm 2D}(k_x,k_y)$ to its exterior, corresponding to the disappearance of the boundary states connecting the two bands from Fig.~\ref{geometry_ykx}(c3) to (e3).

\begin{figure*}
\includegraphics[width=1\linewidth]{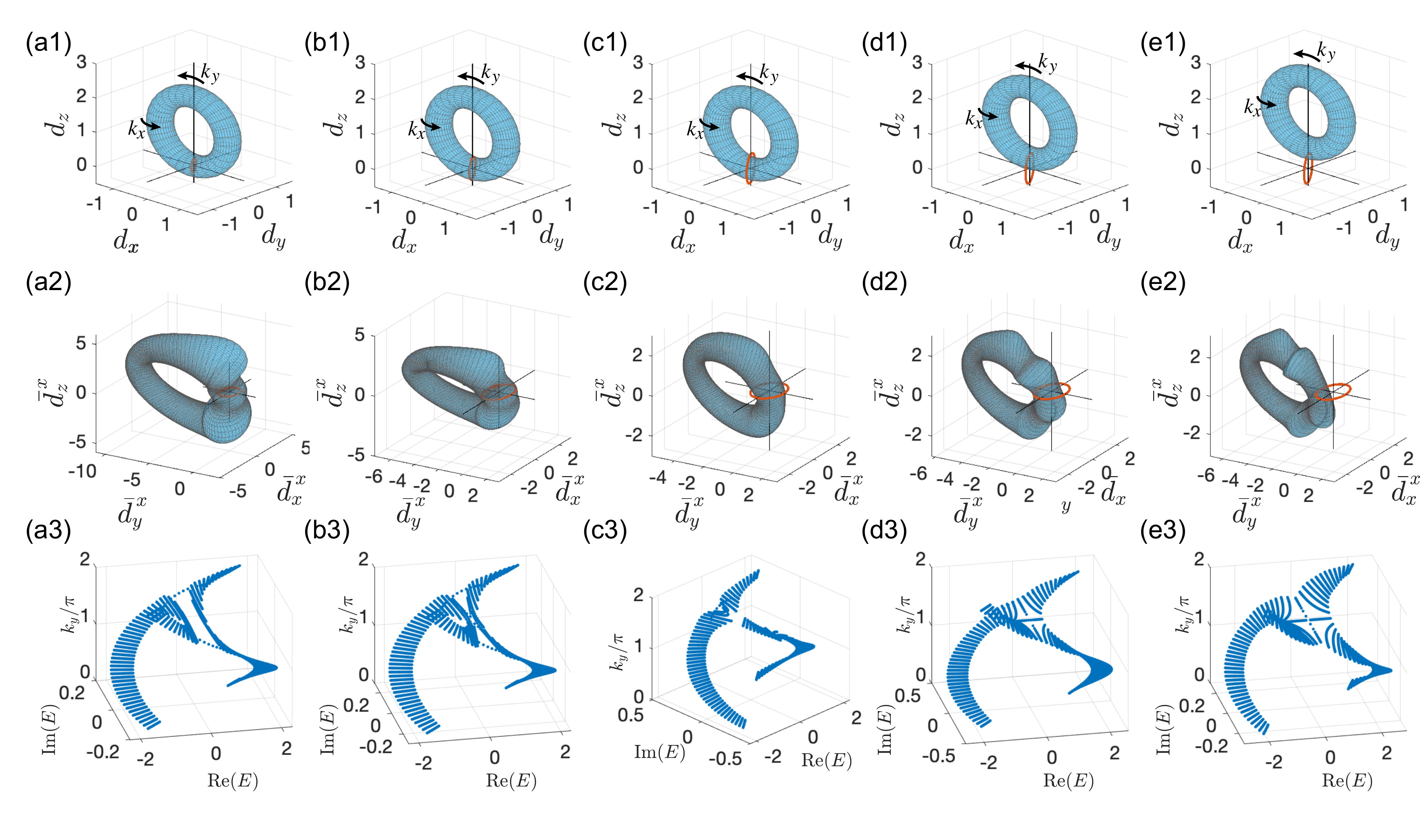}
\caption{Geometric visualization of the 2D effective model of $\bar{H}^x_{\rm 2D}(k_x,k_y)$ after normalizing the SR, \CH{displaying distinct scenarios of SR (a) within, (b,d,e) intersecting, and (c) linking the distorted torus.} The parameters are (a) $g_0=0.2$, $\mu=1$, (b) $g_0=0.3$, $\mu=1$, (c) $g_0=0.4$, $\mu=1$, (d) $g_0=0.4$, $\mu=1.3$, (e) $g_0=0.4$, $\mu=1.6$, with $g_x=g_0\cos\frac{\pi}{6}$ and $g_y=g_0\sin\frac{\pi}{6}$. The other parameters are $t_1=1$, $t_2=0.3$, $g_z=0$.}
\label{geometry_xky}
\end{figure*}

Next, we turn to the complex deformation of \CH{$k_x\rightarrow k_x+i\kappa_x$}, which is associated with with \CH{$x$-OBC} spectrum.
The non-Bloch Hamiltonian
\begin{eqnarray}
\bar{H}^y_{\rm 2D}({\bm k})=H_{\rm 2D}({\bm k}+i\kappa_y)
\end{eqnarray}
yields [see Appendix \ref{app:2D_xOBC}]
\begin{eqnarray}
e^{4\kappa_x}=\frac{(t_1+\mu\sin k_y+g_y)^2+(g_x\cos k_y+g_z\sin k_y)^2}{(t_1+\mu\sin k_y-g_y)^2+(g_x\cos k_y+g_z\sin k_y)^2},\nonumber\\
\end{eqnarray}
\CH{which now also depends on $k_y$, which has been taken as a constant parameter.}
In Fig. \ref{geometry_xky} we illustrate the geometric features of both ${H}_{\rm 2D}(k_x,k_y)$ and $\bar{H}^x_{\rm 2D}(k_x,k_y)$, together with the corresponding $x$-OBC spectrum.
The parameters are chosen as the same as those in Fig. \ref{geometry_ykx}.
In Fig.~\ref{geometry_xky}, the $\bar{\bm d}^x$-torus of the effective $\bar{H}^x_{\rm 2D}(k_x,k_y)$ encloses the SR as a point in panel (a2), links to the SR in panel (c2), and intersect with the SR in panels (b2,d2,e2).
Correspondingly, the spectrum shows three types of band structures, which are (i) the two bands are separated but connected by boundary states [Fig.~\ref{geometry_xky}(a3)], (ii) the two bands are joint into one through $k_y$~\ref{geometry_xky}(c3)], and (iii) the two bands touch each other at certain $k_y$~\ref{geometry_xky}(b3,d3,e3)].
While the $\bar{d}^x({\bm k})$ tori are also distorted by $k_x\rightarrow k_x+i\kappa_x$ substitution and the SR normalization, the resulting $\bar{\bm d}^x$-torus are seen to have the same topological relation to the SR as the one of the ${\bm d}$-torus of the original Hamiltonian.
It is also seen that the $x$-OBC and $y$-OBC spectra show distinguished behaviors with the same parameters, owing to the different NHSEs along $x$ and $y$ directions.

As the BBC is now restored for the $j$-OBC boundary states and the effective Hamiltonian $\bar{H}^j_{\rm 2D}(k_x,k_y)$ with $j=x,y$, we can characterize the topological boundary states under $j$-OBC with a Chern number $C_j$ for $\bar{H}^j_{\rm 2D}(k_x,k_y)$ defined as
\begin{eqnarray}
C_j=\frac{1}{2\pi}\iint d k_x dk_y (\partial_{k_x}A_{j,k_y}-\partial_{k_y}A_{j,k_x}),
\end{eqnarray}
with $A_{j,\alpha}=-{\rm Im}\langle \psi_j^{L} |\partial_\alpha| \psi_j^{R} \rangle$ the non-Hermitian Berry connection, and $\psi_j^{L,R}$ the left/right eigenvector of $\bar{H}^j_{\rm 2D}(k_x,k_y)$ for $E_j=-\sqrt{E_j^2}$. Numerically, we find that $C_y=1$ when there is a pair of boundary modes connecting the two bands [Fig.~\ref{geometry_ykx}(a3-c3)], corresponding to the case where the SR is enclosed by the $\bar{\bm d}^y$-torus [Fig.~\ref{geometry_ykx}(a2-c2)]; and $C_y=0$ when no boundary mode connects the two bands [Fig.~\ref{geometry_ykx}(e3)], corresponding to the case where the SR is outside the $\bar{\bm d}^y$-torus [Fig.~\ref{geometry_ykx}(e2)]. Similar to the Chern number defined for Hermitian systems, $C_y$ is also ill-defined when the SR and the $\bar{\bm d}^y$-torus intersect [Fig.~\ref{geometry_ykx}(d2)], where the two bands touch each other [Fig.~\ref{geometry_ykx}(d3)].
Similarly, the other Chern number \CH{$C_x=1$ in the case of Fig.~\ref{geometry_xky}(a2,a3),  $C_x=0$ in the case of Fig.~\ref{geometry_xky}(c2,c3) (even when the SR and the $\bar{\bm d}^x$-torus are linked), and is ill-defined for the rest cases we have discussed.}


\section{Summary}\label{sec:sum}
\CH{We have introduced the Singularity Ring formalism for visualizing the geometric and topological properties of general two-component non-Hermtian systems. By explicitly expressing band touching singularities as a ring instead of a point in a 3D vector space, it illuminates the physical implications of the reduced codimension of generic singularities due to non-Hermiticity.  }

\CH{The most immediate observation is that the band vorticity can be interpreted as the linking number between the SR and a ${\bm d(k)}$-loop defining the Hermitian part of the Hamiltonian, which can hence be measured through the Berry phase. In 2D, the ${\bm d(k)}$-loop is promoted to a ${\bm d}$-torus which can intersect, link or enclose the SR in various interesting manners. Besides the winding structure of corresponding PBC bands, their geometric configuration can also reveal the Chern number signifying the number of protected chiral edge modes.}

\CH{The SR formalism proves even more useful when classifying symmetry-protected systems, where symmetry constraints translate to easily visualized restrictions on the possible configurations of the ${\bm d(k)}$-loop and the SR, as summarized in Table I. For CS classes, the $Z$ topological number is just the winding number of the ${\bm d(k)}$-loop about the SR in its plane, while for SLS, PHS and PHS$^\dagger$ classes, the $Z$ and $Z_2$ number can be counted from how the ${\bm d(k)}$ intersects particular planes.}

\CH{Beyond the 1D and 2D representative variants of the SSH model for the four symmetries considered, our SR approach is amenable to the most general non-Hermitian two-component models. Models expressed in different bases, or those with generic non-Hermitian terms, can all be put on equal footing via the normalization procedure described in Sect.~II B, where the non-Hermitian singularity is always represented as a unit ring. Importantly, this flexibility accommodates the use of a non-Bloch basis for studying systems afflicted by the non-Hermitian skin effect.}


Indeed, our work generalizes the Bloch sphere interpretation for Hermitian topological systems by establishing connections between the topological and the geometric properties of two-component non-Hermitian Hamiltonians. The geometric SR picture characterizes not just the vorticity of the PBC bands, but also full topological information like the Berry phase and Chern number.
For multi-band systems whose Hamiltonians contain only anticommuting terms satisfying a Clifford algebra, an analogous SR picture holds, although the vector space is no longer limited to 3D.  The singularity manifold will have a dimension of $D-2$, with $D$ the dimension of the vector space.
Such higher-dimensional objects shall lead to more complicated but fascinating geometric and topological features. In more general cases, however, where the Hamiltonian cannot be mapped onto a vector space with an orthogonal basis, the relation between geometric features and topological properties have yet to be explored.



\begin{acknowledgments}
J.G. acknowledges
fund support by the Singapore NRF grant No. NRF-NRFI2017- 04 (WBS No. R-144-000-378-281).
\end{acknowledgments}

\appendix

\section{Detecting the linking number through a dynamical process}\label{app:dynamical}
In this appendix, we discuss a dynamical process to extract the Berry phase $\gamma_{\rm sum}$ for the model
\begin{eqnarray}
h(k)&=&(t_1+t_2\cos{k}+t'\cos{3k}+ig_x)\sigma_x\nonumber\\
&&+(t_2\sin{k}+t'\sin{3k}+ig_y)\sigma_y+ig_z\sigma_z]\label{eq.app:hk}
\end{eqnarray}
discussed in Section~\ref{sec:SR_Berry}.
When the system has an odd linking number and the two bands joint into one loop \CH{(half-integer vorticity)}, the two eigenmodes will interchange after $k$ varies a period of $2\pi$, and a given state returns to itself only after two periods. In this process the state evolves through both bands, and shall acquire the Berry phase $\gamma_{\rm sum}=\pi$. On the contrary, in the cases with an even linking number \CH{(integer vorticity)}, a given eigenmode shall only go through the same band twice after $k$ varies from $0$ to $2\pi$, and thus cannot capture $\gamma_{\rm sum}$.

To see this, we can consider an adiabatic dynamical process of a two-level single-particle system,
and take $k=\omega t$ to simulate a 1D model.
However, an adiabatic evolution is impossible here as the joining of two bands requires complex eigen-energies, which shall induce gain and loss and lead to attenuation or instabilities along the evolution.
To avoid this, we consider an alternative Hamiltonian
\begin{eqnarray}
h'(\omega t)=e^{-i\phi(\omega t)}h(\omega t),\label{h_dynamical}
\end{eqnarray}
with $\phi(\omega t)$ the phase angle of the eigen-energies at $\omega t$.
This $h'(\omega t)$ has the same eigenmodes as the original Hamiltonian, thus $\gamma_{\rm sum}$ shall be the same. On the other hand, it only has real eigen-energies, hence the $PT$ symmetry is restored effectively, and an adiabatic process becomes possible~\cite{Gong2010PT}.
Preparing an initial state $\psi(0)$ as one of the eigenmode of $h'(0)$, say $\psi(0)=u_+^R(0)$, the final state at $t=T$ is given by
\begin{eqnarray}
\psi(\omega T)=e^{-i\int_0^{T}h'(\omega t)dt}u_+^R(0)\label{eq:evolution}.
\end{eqnarray}
If the two bands are \CH{connected}, $\psi(\omega T=4\pi)$ shall agree with $u(0)$ up to only a phase difference. As the two bands are associated with $E_+(k)=E_-(k)$, we can also expect the dynamical phase cancel out itself when the system varies two periods, leaving only the Berry phase $\gamma_{\rm sum}=\pi$.
On the other hand, if the two bands are separated, such a process only goes through the same band twice, obtaining a Berry phase of $2\gamma_{+}$.
However, now the dynamical phase cannot cancel out itself, and shall overwhelm the Berry phase obtained from the evolution.

\begin{figure}
\includegraphics[width=0.99\linewidth]{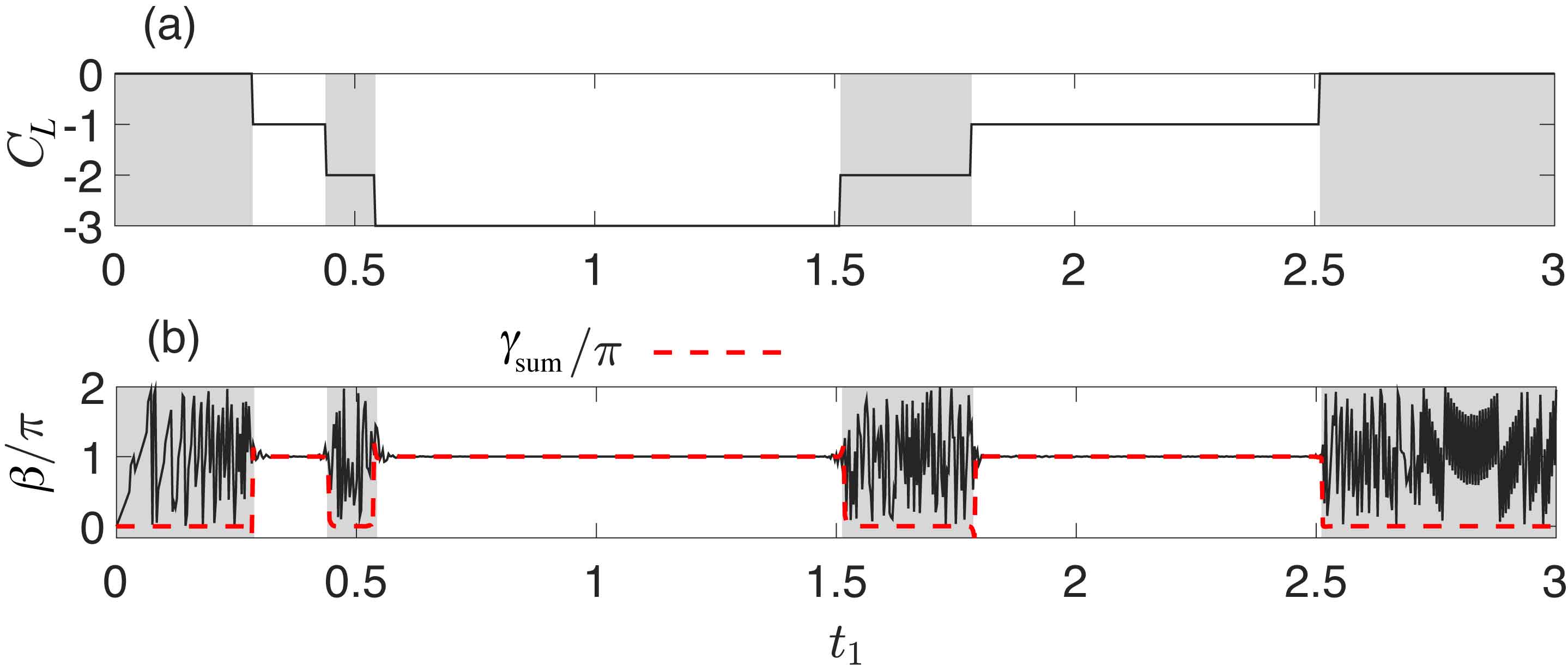}
\caption{(a) The linking number $C_L$ as a function of $t_1$ for the Hamiltonian (\ref{eq.app:hk}).
In panels (b), black line shows the phase angle $\beta$ defined in Eq. (\ref{inner}), and red dash line show the summed Berry phase $\gamma_{\rm sum}$. \CH{Both agree at odd $C_L$ i.e. nontrivial vorticity, as explained in the main text.} Other parameters are $t_2=0.5$, $t'=1$, $g_x=g_z=0.3$, $g_y=1$.}
\label{fig:app1}
\end{figure}

Numerically, the time evolution of Eq. (\ref{eq:evolution}) can be simulated by
\begin{eqnarray}
\psi(\omega T)=\prod_{m=1}^M e^{-ih'(\omega m\Delta t)}u_+^R(0)
\end{eqnarray}
with $\Delta t= T/M$. In our simulation, we find that the final state $\psi(\omega T)$ at $\omega T=4\pi$ always recovers the initial one with a complex coefficient, indicating an adiabatic following of the system. Therefore, defining
\begin{eqnarray}
Ae^{i\beta}=\langle \psi(0) | \psi(\omega T=4\pi)\rangle,\label{inner}
\end{eqnarray}
we shall obtain $\beta=\gamma_{\rm sum}=\pi$ when the two bands are joint, and a random $\beta$ otherwise, due to the contribution of a dynamical phase. On the other hand, the quantity $A$ is generally not one, as an unitary evolution is not guaranteed in non-Hermitian system, even in an adiabatic process.
In Fig.~\ref{fig:app1}(b) we illustrate the value of $\beta$, which is consistent with our analysis, and the results of the Berry phase $\gamma_{\rm sum}$ (red dashed line) and the linking number [Fig.~\ref{fig:app1}(a)].


In this dynamical process, the adiabatic following is made possible by the real spectrum of Hamiltonian (\ref{h_dynamical}), which is difficult to be realized in a lattice system.
Nevertheless, we can map the momentum space of a lattice system to a controllable parameter space of a two-level single-particle system, e.g. a superconducting transmon qubit~\cite{tan2018demonstration}.
Thus Hamiltonian (\ref{h_dynamical}) can be realized by tuning the parameters correspondingly, and the Berry phase can be detected through
the Ramsey fringe interference technique~\cite{tan2018demonstration,leek2007observation,tan2014demonstration}.
The non-Hermiticity of $h(\omega t)$ corresponds to tunable gain and loss on the two levels of the qubit, \blue{which can be induced by coupling the qubit to a controllable environment.}

\section{Derivation of $\kappa$ for different systems}
\subsection{Sublattice symmetry (SLS)}\label{app:SLS}
The SL-symmetric SSH model is described by
\begin{eqnarray}
{H}(k)=({\bm d}+i{\bm g})\cdot{\bm \sigma}
\end{eqnarray}
with
\begin{eqnarray}
{\bm d}&=&(t_1+t_2\cos k, t_2\sin k,0),\nonumber\\
{\bm g}&=&(g_x,g_y,0).
\end{eqnarray}
By direct substitution of $k\rightarrow k+i\kappa$, the non-Bloch Hamiltonian is given by
\begin{eqnarray}
\bar{H}(k)=(\bar{\bm d}+i\bar{\bm g})\cdot{\bm \sigma},
\end{eqnarray}
with
\begin{eqnarray}
\bar{d}_x(k)&=&t_1+t_2\frac{e^{-\kappa}+e^{\kappa}}{2}\cos k,\nonumber\\
\bar{d}_y(k)&=&t_2\frac{e^{-\kappa}+e^{\kappa}}{2}\sin k,\nonumber\\
\bar{g}_x(k)&=&g_x+t_2\frac{e^{-\kappa}-e^{\kappa}}{2}\sin k,\nonumber\\
\bar{g}_y(k)&=&g_y-t_2\frac{e^{-\kappa}-e^{\kappa}}{2}\cos k,\nonumber\\
\bar{d}_z(k)&=&\bar{g}_z(k)=0.
\end{eqnarray}
and eigen-energies given by
\begin{eqnarray}
\bar{E}_{\pm}^2&=&f_0+f_1\cos k+f_2\sin k\nonumber\\
&&+i(f_3\cos k+f_4\sin k),
\end{eqnarray}
where
\begin{eqnarray}
f_0&=&t_1^2+t_2^2-g_x^2-g_y^2,\nonumber\\
&&+2it_1g_x\nonumber\\
f_1&=&t_1t_2(e^{-\kappa}+e^{\kappa})+g_yt_2(e^{-\kappa}-e^{\kappa}),\nonumber\\
f_2&=&-g_xt_2(e^{-\kappa}-e^{\kappa}),~~f_3=g_xt_2(e^{-\kappa}+e^{\kappa}),\nonumber\\
f_4&=&t_1t_2(e^{-\kappa}-e^{\kappa})+g_yt_2(e^{-\kappa}+e^{\kappa}),\nonumber
\end{eqnarray}
\CH{all dependent on $\kappa$.  By requiring that the spectrum forms a doubly degenerate arc i.e. $\bar{E}_\pm(\alpha+k)=\bar{E}_\pm(\alpha-k)$ for some offset $\alpha$ and $k$. A special $\kappa$ solution for removing NHSE is found from
 $\tan \alpha=f_2/f_1$ when $f_1f_4=f_2f_3$ is satisfied. Simplifying, we obtain }
\begin{eqnarray}
e^{4\kappa}=\frac{t_1^2+g_y^2+2t_1g_y+g_x^2}{t_1^2+g_y^2-2t_1g_y+g_x^2}.
\end{eqnarray}

\subsection{Conjugated particle-hole symmetry  (PHS$^\dagger$)}\label{app:PHSdag}
The PH$^\dagger$-symmetric SSH model is described by
\begin{eqnarray}
{H}(k)=({\bm d}+i{\bm g})\cdot{\bm \sigma}
\end{eqnarray}
with
\begin{eqnarray}
{\bm d}=(t_1+t_2\cos k, t_2\sin k,0),\nonumber\\
{\bm g}=(0,g_y,g_z).
\end{eqnarray}
The \CH{non-Bloch} Hamiltonian $\bar{H}(k)= H(k\rightarrow k+i\kappa)$ takes the form
\begin{eqnarray}
\bar{d}_x(k)&=&t_1+t_2\frac{e^{-\kappa}+e^{\kappa}}{2}\cos k,\nonumber\\
\bar{d}_y(k)&=&t_2\frac{e^{-\kappa}+e^{\kappa}}{2}\sin k,\nonumber\\
\bar{g}_x(k)&=&t_2\frac{e^{-\kappa}-e^{\kappa}}{2}\sin k,\nonumber\\
\bar{g}_y(k)&=&g_y-t_2\frac{e^{-\kappa}-e^{\kappa}}{2}\cos k,\nonumber\\
\bar{g}_z(k)&=&g_z,~~\bar{d}_z(k)=0.\label{effectiveH_SSH3}
\end{eqnarray}
Its eigen-energies are given by
\begin{eqnarray}
\bar{E}_{\pm}^2=f_0+f_1\cos k+i(f_4\sin k),
\end{eqnarray}
with
\begin{eqnarray}
f_0&=&t_1^2+t_2^2-g_y^2-g_z^2,\nonumber\\
f_1&=&t_1t_2(e^{-\kappa}+e^{\kappa})+g_yt_2(e^{-\kappa}-e^{\kappa}),\nonumber\\
f_4&=&t_1t_2(e^{-\kappa}-e^{\kappa})+g_yt_2(e^{-\kappa}+e^{\kappa}).\nonumber
\end{eqnarray}

Similar to the \blue{PH$^\dagger$-symmetric SSH model}, the NHSE is removed when $f_1f_4=0$ where $\bar{E}_{\pm}(k)=\bar{E}_{\pm}(-k)$, with $\kappa$ given by
\begin{eqnarray}
e^{4\kappa}=\frac{t_1^2+g_y^2+2t_1g_y}{t_1^2+g_y^2-2t_1g_y}.\label{kappa_SSH3}
\end{eqnarray}

\subsection{2D system with $y$-OBC}\label{app:2D_yOBC}
For the 2D system with
\begin{eqnarray}
d_x(k_x,k_y)&=&(t_1+t_2\cos{k_x})\cos{k_y}\nonumber\\
d_y(k_x,k_y)&=&t_2\sin{k_x}\nonumber\\
d_z(k_x,k_y)&=&(t_1+t_2\cos{k_x})\sin{k_y}+\mu,
\end{eqnarray}
the eigen-energies of the effective Hamiltonian $\bar{H}^y_{\rm 2D}({\bm k})=H_{\rm 2D}({\bm k}+i\kappa_y)$ are given by
\begin{eqnarray}
E_y^2({\bm k}+i\kappa_y)&=&f_0^y+f_1^y\cos k_y+f_2^y\sin k_y\nonumber\\
&&+i(f_3^y\cos k_y+f_4^y\sin k_y),
\end{eqnarray}
with
\begin{eqnarray}
f_0^y&=&t_1^2+2t_1t_2\cos{k_x}+t_2^2+\mu^2-g_x^2-g_y^2-g_z^2\nonumber\\
&&+2i(g_yt_2\sin k_x+g_z\mu)\nonumber\\
f_1^y&=&g_z(t_1+t_2\cos k_x)(e^{-\kappa_y}-e^{\kappa_y}),\nonumber\\
f_2^y&=&\mu(t_1+t_2\cos k_x)(e^{-\kappa_y}+e^{\kappa_y})\nonumber\\
&&-g_x(t_1+t_2\cos k_x)(e^{-\kappa_y}-e^{\kappa_y}),\nonumber\\
f_3^y&=&-\mu(t_1+t_2\cos k_x)(e^{-\kappa_y}-e^{\kappa_y})\nonumber\\
&&+g_x(t_1+t_2\cos k_x)(e^{-\kappa_y}+e^{\kappa_y}),\nonumber\\
f_4^y&=&g_z(t_1+t_2\cos k_x)(e^{-\kappa_y}+e^{\kappa_y}).\nonumber
\end{eqnarray}
Requiring $f_1^yf_4^y=f_2^yf_3^y$ \CH{as in the 1D SLS case} leads to
\begin{eqnarray}
e^{4\kappa_y}=\frac{\mu^2+\gamma_x^2+\gamma_z^2-2\mu\gamma_x}{\mu^2+\gamma_x^2+\gamma_z^2+2\mu\gamma_x},\nonumber\\
\end{eqnarray}
which \CH{does not depend on $k_x$}.

\subsection{2D system with $x$-OBC}\label{app:2D_xOBC}
For the complex deformation of \CH{$k_x\rightarrow k_x+i\kappa_x$} associated with topological boundary states with \CH{$x$-OBCs},
the eigen-energies of the non-Bloch Hamiltonian
\begin{eqnarray}
\bar{H}^x_{\rm 2D}({\bm k})=H_{\rm 2D}({\bm k}+i\kappa_x),
\end{eqnarray}
are given by\begin{eqnarray}
E_x^2({\bm k}+i\kappa_x)&=&f_0^x+f_1^x\cos k_x+f_2^x\sin k_x\nonumber\\
&&+i(f_3^x\cos k_x+f_4^x\sin k_x),
\end{eqnarray}
with
\begin{eqnarray}
f_0^x&=&t_1^2+t_2^2+\mu^2+2\mu t_1\sin k_y-g_x^2-g_y^2-g_z^2\nonumber\\
&&+2i[t_1(g_x\cos k_y+g_z\sin k_y)+g_z\mu]\nonumber\\
f_1^x&=&t_2(t_1+\mu\sin k_y)(e^{-\kappa_x}+e^{\kappa_x})\nonumber\\
&&+g_yt_2(e^{-\kappa_x}-e^{\kappa_x}),\nonumber
\end{eqnarray}
\begin{eqnarray}
f_2^x&=&-t_2(g_x\cos k_y+g_z\sin k_y)(e^{-\kappa_x}-e^{\kappa_x}),\nonumber\\
f_3^x&=&t_2(g_x\cos k_y+g_z\sin k_y)(e^{-\kappa_x}+e^{\kappa_x}),\nonumber\\
f_4^x&=&t_2(t_1+\mu\sin k_y)(e^{-\kappa_x}-e^{\kappa_x})\nonumber\\
&&+g_yt_2(e^{-\kappa_x}+e^{\kappa_x}).\nonumber
\end{eqnarray}
Similar to the $y$-OBC case above and the 1D model with SLS, the NHSE is eliminated when $f_1^xf_4^x=f_2^xf_3^x$, which leads to
\begin{eqnarray}
e^{4\kappa_x}=\frac{(t_1+\mu\sin k_y+g_y)^2+(g_x\cos k_y+g_z\sin k_y)^2}{(t_1+\mu\sin k_y-g_y)^2+(g_x\cos k_y+g_z\sin k_y)^2},\nonumber\\
\end{eqnarray}
\CH{which now also depends on $k_y$, which has been taken as a constant parameter.}

\bibliography{references}

\clearpage

\end{document}